\documentclass[twocolumn, trackchanges]{aastex62}
\shorttitle{Torus Covering Factor Of \swift/BAT AGN}
\shortauthors{Ichikawa et al.}

\usepackage{amsmath}
\usepackage{color}
\usepackage{url}
\usepackage{ulem}
\usepackage{multirow}
\usepackage{longtable}
\usepackage{graphicx}
\usepackage{epstopdf}

\newcommand{\swift}{\textit{Swift}}
\newcommand{\wise}{\textit{WISE}}
\newcommand{\akari}{\textit{AKARI}}
\newcommand{\iras}{\textit{IRAS}}
\newcommand{\herschel}{\textit{Herschel}}

\newcommand{\lagntwelve}{L^{\rm (AGN)}_{12~\mu{\rm m}}}
\newcommand{\lagnmir}{L^{\rm (AGN)}_{{\rm MIR}}}
\newcommand{\lagnir}{L^{\rm (AGN)}_{{\rm IR}}}
\newcommand{\ltorus}{L^{\rm (AGN;1-1000{\mu}m)}_{{\rm IR}}}
\newcommand{\lbol}{L^{\rm (AGN)}_{\rm bol}}

\newcommand{\fagntwelve}{f_{\rm AGN}^{(12 \mu{\rm m})}}
\newcommand{\fagnmir}{f_{\rm AGN}^{(\rm MIR)}}
\newcommand{\fagnir}{f_{\rm AGN}^{(\rm IR)}}

\newcommand{\cfdust}{C_{\rm T}({\rm dust})}
\newcommand{\cfgasdust}{C_{\rm T}({\rm gas+dust})}
\newcommand{\cfdustsim}{C_{\rm T}({\rm dust;sim})}

\begin{document}


\title{BAT AGN Spectroscopic Survey -- XI. The Covering Factor of
Dust and Gas in \swift/BAT Active Galactic Nuclei}


\correspondingauthor{Kohei Ichikawa}
\email{k.ichikawa@astr.tohoku.ac.jp}

\author[0000-0002-4377-903X]{Kohei Ichikawa}
\affil{
Department of Astronomy, Columbia University, 550 West 120th Street, New York, NY 10027, USA}
\affil{
Department of Physics and Astronomy, University of Texas at San Antonio, One UTSA Circle, San Antonio, TX 78249, USA}
\affil{
Frontier Research Institute for Interdisciplinary Sciences, Tohoku University, Sendai,
Miyagi 980-8578, Japan}
\affil{
Astronomical Institute, Tohoku University, Aramaki, Aoba-ku, Sendai, Miyagi 980-8578, Japan}

\author{Claudio Ricci}
\affiliation{
N\'ucleo de Astronom\'ia de la Facultad de Ingenier\'ia, Universidad Diego Portales, Av. Ej\'ercito Libertador 441, Santiago, Chile}
\affiliation{
Kavli Institute for Astronomy and Astrophysics, Peking University, Beijing 100871, China
}
\affiliation{
Chinese Academy of Sciences South America Center for Astronomy, Camino El Observatorio 1515, Las Condes, Santiago, Chile}

\author{Yoshihiro Ueda}
\affiliation{
Department of Astronomy, Kyoto University, Oiwake-cho, Sakyo-ku, Kyoto 606-8502, Japan
}

\author{Franz E. Bauer}
\if0
\affiliation{
Instituto de Astrof{\'{\i}}sica and Centro de Astroingenier{\'{\i}}a, Facultad de F{\'{i}}sica, Pontificia Universidad Cat{\'{o}}lica de Chile, Casilla 306, Santiago 22, Chile
}
\fi
\affiliation{
Institute of Astrophysics, Pontificia Universidad Catolica de Chile, Avenida Vicua Mackenna 4860, 7820436, Chile}
\affiliation{
Millennium Institute of Astrophysics (MAS), Nuncio Monse{\~{n}}or S{\'{o}}tero Sanz 100, Providencia, Santiago, Chile
}
\affiliation{
Space Science Institute, 4750 Walnut Street, Suite 205, Boulder, Colorado 80301, USA
}

\author{Taiki Kawamuro}
\altaffiliation{JSPS fellow}
\affiliation{
National Astronomical Observatory of Japan, 2-21-1 Osawa, Mitaka, Tokyo 181-8588, Japan}

\author{Michael J. Koss}
\affiliation{
Eureka Scientific, 2452 Delmer Street Suite 100, Oakland, CA 94602-3017, USA
}

\author{Kyuseok Oh}
\altaffiliation{JSPS fellow}
\affiliation{
Department of Astronomy, Kyoto University, Oiwake-cho, Sakyo-ku, Kyoto 606-8502, Japan
}

\author{David J. Rosario}
\affiliation{
Department of Physics, Durham University, South Road, DH1 3LE Durham, UK
}

\author{T. Taro Shimizu}
\affiliation{
Max-Planck-Institut f\"{u}r extraterrestrische Physik, Postfach 1312, 85741, Garching, Germany
}

\author{Marko Stalevski}
\affiliation{
Astronomical Observatory, Volgina 7, 11060 Belgrade, Serbia
}
\affiliation{
Sterrenkundig Observatorium, Universiteit Gent, Krijgslaan 281-S9, Gent B-9000, Belgium
}

\author{Lindsay Fuller}
\affiliation{
Department of Physics and Astronomy, University of Texas at San Antonio, One UTSA Circle, San Antonio, TX 78249, USA}

\author{Christopher Packham}
\affiliation{
Department of Physics and Astronomy, University of Texas at San Antonio, One UTSA Circle, San Antonio, TX 78249, USA}
\affiliation{
National Astronomical Observatory of Japan, 2-21-1 Osawa, Mitaka, Tokyo 181-8588, Japan}

\author{Benny Trakhtenbrot}
\affiliation{
Department of Physics, ETH Zurich, Wolfgang-Pauli-Strasse 27, CH-8093 Zurich, Switzerland
}
\affiliation{
School of Physics and Astronomy, Tel Aviv University, Tel Aviv 69978, Israel
}



\begin{abstract}

We quantify the luminosity contribution of active galactic nuclei (AGN) to
the 12~$\mu$m, mid-infrared (MIR; 5-38~$\mu$m),
and the total IR (5--1000~$\mu$m) emission
 in the local AGN detected in the all-sky 70-month
\swift/Burst Alert Telescope (BAT) ultra hard X-ray survey.
We decompose the IR spectral energy distributions (SEDs) of 587 objects into
AGN and starburst components using AGN torus and star-forming galaxy templates.
This enables us to recover the AGN torus emission also for low-luminosity end, down to
 $\log (L_{14-150}/{\rm erg}~{\rm s}^{-1})\simeq 41$, 
which typically have significant host galaxy contamination.
We find that the luminosity contribution of the AGN to the 12~$\mu$m, 
the MIR, and the total IR band
is an increasing function of the 14--150~keV luminosity.
We also find that for the most extreme cases, the IR pure-AGN
emission from the torus can extend up to 90~$\mu$m.
The obtained total IR AGN luminosity through the IR SED decomposition
enables us to estimate the fraction of the sky obscured by dust, i.e., the dust covering
factor. 
We demonstrate that the median of the dust covering factor
is always smaller than that of the X-ray obscuration fraction
  above the AGN bolometric luminosity of
  $\log (\lbol/{\rm erg}~{\rm s}^{-1}) \simeq 42.5$. 
  Considering that X-ray obscuration fraction
  is equivalent to the covering factor coming from both the dust and gas,
 it indicates that an additional neutral gas component, 
 along with the dusty torus, is responsible for the absorption of X-ray emission.
\end{abstract}


%% subject header.  Objects should be in the appropriate "individual"
%% headers (e.g. quasars: individual, stars: individual, etc.) with the
%% additional provision that the total number of headers, including each
%% individual object, not exceed six.  The \objectname{} macro, and its
%% alias \object{}, is used to mark each object.  The macro takes the object
%% name as its primary argument.  This name will appear in the paper
%% and serve as the link's anchor in the electronic edition if the name
%% is recognized by the data centers.  The macro also takes an optional
%% argument in parentheses in cases where the data center identification
%% differs from what is to be printed in the paper.



%%%%%%%%%%%%%%INTRODUCTION%%%%%%%%%%%%%%%%%%
%%%%%%%%%%%%%%INTRODUCTION%%%%%%%%%%%%%%%%%%

\section{INTRODUCTION}
%AGN is important candidate to know the cosmic evolution of SMBH and the host galaxies.
One of the fundamental open questions of extragalactic astrophysics is how supermassive black holes (SMBHs) and their host galaxies
co-evolve \citep[e.g.,][]{ale12}.
Active galactic nuclei (AGN) are the best targets
 to understand this process of coevolution, because they are in the stage
where the mass accretion onto SMBHs occurs by releasing large amounts of radiation \citep[e.g.,][]{yu02,mar04},
until they reach their achievable maximum mass of $M_{\rm BH} \simeq 10^{10.5} M_{\odot}$ \citep{net03,mcl04,tra14,jun15,ina16,ich17b}.

%X-ray is the most suited energy band to construct a census of AGN
Ultra-hard ($E>10$~keV) X-ray observations are one of the most reliable methods for identifying AGN.
Thanks to the combination of (1) a strong penetration power up to $\log (N_{\rm H}/{\rm cm}^{-2}) \simeq 24$
\citep[e.g.,][]{ric15}
and (2) the high contrast over stellar X-ray emission \citep[e.g.,][]{min12},
the ultra-hard X-ray surveys allow the potential for an unbiased census
of AGN up to Compton-thick levels \citep[e.g.,][]{kos16}.
Among the recent available surveys, 
\swift/BAT provides the most sensitive X-ray survey of the whole sky
 in the 14--195 keV range, reaching a flux level of (1.0--1.3)$\times 10^{-11}$~erg~s$^{-1}$~cm$^{-2}$ in the first 70 months of 
 operations \citep{bau13}, and
 to the deeper flux of
 (7.2--8.4)$\times 10^{-12}$~erg~s$^{-1}$~cm$^{-2}$
 in the recently updated 105-month catalog \citep{oh18}.

%IR is another method to know AGN activity
Infrared (IR) observations also provide an effective method
to study AGN because the central engine of AGN is expected to be surrounded
by a dusty ``torus'' \citep{kro86} which is heated by the AGN
and re-emits thermally in the mid-IR (MIR)
\citep[e.g.,][]{gan09,asm15,ich12,ich17}.
%IR Soltan's argument
A recent upward revision of black hole scaling relations \citep{kor13} indicates that the local mass density in black holes
should be higher, suggesting that 
a larger population of heavily obscured AGN gas and dust
 is required to fill the mass gap of the revised local black hole mass density \citep[e.g.,][]{nov13,com15}. 
 These populations contribute significantly to the infrared background \citep[e.g.,][]{mur11, del14}, especially in the MIR band \citep{ris02}.
 %Removing the contamination from SB is necessary
However, since star formation from the host galaxy sometimes contaminates
the MIR emission, especially for low-luminosity AGN with $L_{\rm 14-150}<10^{43}$~erg~s$^{-1}$ \citep[e.g.,][]{ich17}, 
and the torus is too compact \citep[$<10$~pc; eg.,][]{jaf04} to be fully resolved
\citep[e.g.,][]{gar16a,ima18},
 the precise estimation of AGN thermal activity is not straightforward.
 
 %Several methods to remove host components
 %High spatial resolution obs. or IR interferometries
Fortunately, several methods have been proposed to isolate the torus emission from the starburst component.
 One of them is to use high spatial resolution ($\sim$0\farcs3--0\farcs7) 
 MIR observations to resolve the starburst emission of the host galaxies down to 10~pc scales \citep{pac05, rad08, hon10, ram11, alo11, gon13, asm14, ich15,alo16,mar17}.
 In addition, the advent of IR interferometry observations, with their exquisite resolving power (with baselines up to $130$~m),
has spatially resolved the dusty nuclear region and shown that their outer radii in the MIR 
 are typically several pc \citep[e.g.,][]{jaf04,rab09,bur13}.
 Notably, some show the polar elongated dust emission
 suggestive of the dusty outflows \citep{hon12,hon13,tri14,lop16}.
 However, because of the limited sensitivity and the spatial resolution of current telescopes
 \citep[see a recent review of ][]{bur16}, these two methods are available
 only for a few tens of bright sources located in the very local Universe ($z<0.01$).
 
 %SED decomposition
 Another possible approach is to separate the spectral emission of the AGN
 and the starburst (SB) component. Multiple decomposition methods have been applied
 to MIR spectra, mainly using aromatic features as a proxy of star formation (e.g., \citealp{tra01, lut04, saj07, alo12, ich14, her15,kir15, sym16}), to 
 broadband IR spectral energy distributions \citep[SEDs, e.g.,][]{hat08, dac08, xu15,
 lyu16, lyu17,shi17}, and to the combination of both spectra and SEDs \citep[e.g.,][]{mul11}.
 The advantage of the SED decomposition is that 
 it is less affected by the differing spatial resolutions inherent in aperture photometry,
 and can be applied to high-$z$ sources \citep[e.g.,][]{sta15, mat16, lyu16} and/or to
 large ($N>100$) samples, for which high-spatial resolution MIR imaging
 and spectroscopy would require significant amounts of large diameter ($>8$~m) telescope time.

%Our studies: SED decomposition by using the large IR data
In this paper, we decompose the IR SEDs
of ultra-hard X-ray selected \swift/BAT 70-month AGN catalog \citep{bau13} into AGN and host galaxy components.
Thanks to the intensive follow-up observations by the
BAT AGN Spectroscopic Survey\footnote{\url{www.bass-survey.com}}
\citep[BASS;][]{kos17,lam17,ric16},
we are able to obtain the reliable information on
the gas column density ($N_{\rm H}$), 
absorption corrected 14--150~keV X-ray luminosity ($L_{14-150}$), 
and black hole mass ($M_{\rm BH}$) of the sample.

The main goal of this work is to quantitatively assess the AGN contribution 
to 12~$\mu$m, MIR (5--38~$\mu$m) band, and total IR (5--1000~$\mu$m) 
band down to $\log (L_{14-150}/{\rm erg}~{\rm s}^{-1})\simeq 41$ 
in order to investigate 
1) the MIR/X-ray luminosity correlation and 2) 
the dust covering factor of the torus, 
avoiding issues related to host galaxy contamination.
Throughout the paper, we adopt
standard cosmological parameters ($H_{0}=70.0$~km~s$^{-1}$~Mpc$^{-1}$, 
$\Omega_{\rm M}=0.3$, and $\Omega_{\Lambda}=0.7$).

%------------------------------------------fig:zdist-------------------------------------%
\begin{figure}
\begin{center}
\includegraphics[width=\linewidth]{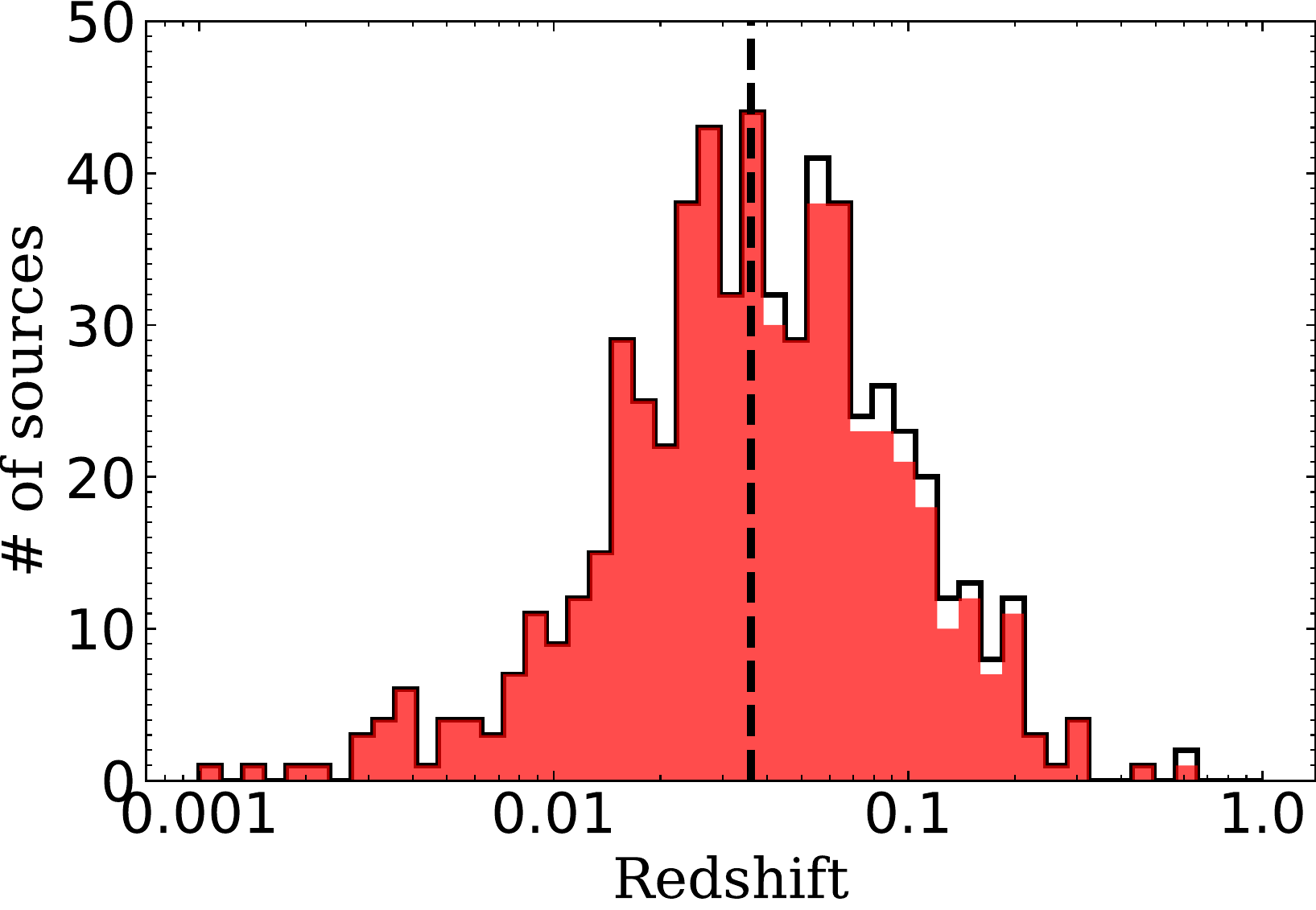}~
\caption{
Redshift distribution of AGN in the \textit{Swift}/BAT
 70-month catalog at galactic latitude of $|b|>10^{\circ}$
 (black solid line: 606 objects; see also \cite{ich17}) and 
of those used in this study (red color area: 587 objects).
The vertical dashed line represents the median of
the redshift ($\left< z \right>=0.037$) in our sample.
}\label{fig:zdist}
\end{center}
\end{figure}
%------------------------------------------fig:zdist-------------------------------------%

%------------------------------------------fig:SEDexample-------------------------------------%
\begin{figure*}
\begin{center}
\includegraphics[width=0.33\linewidth]{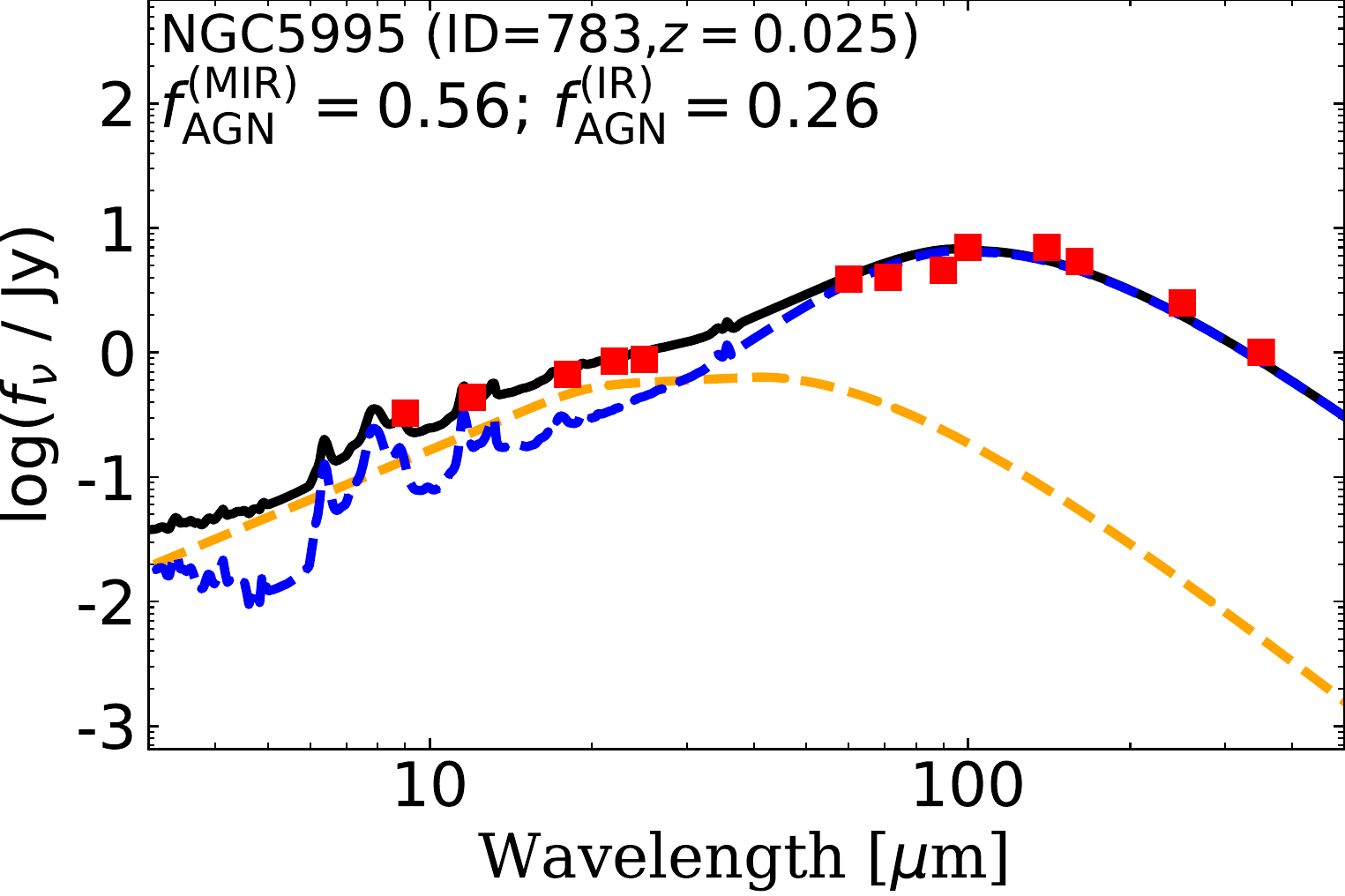}~
\includegraphics[width=0.33\linewidth]{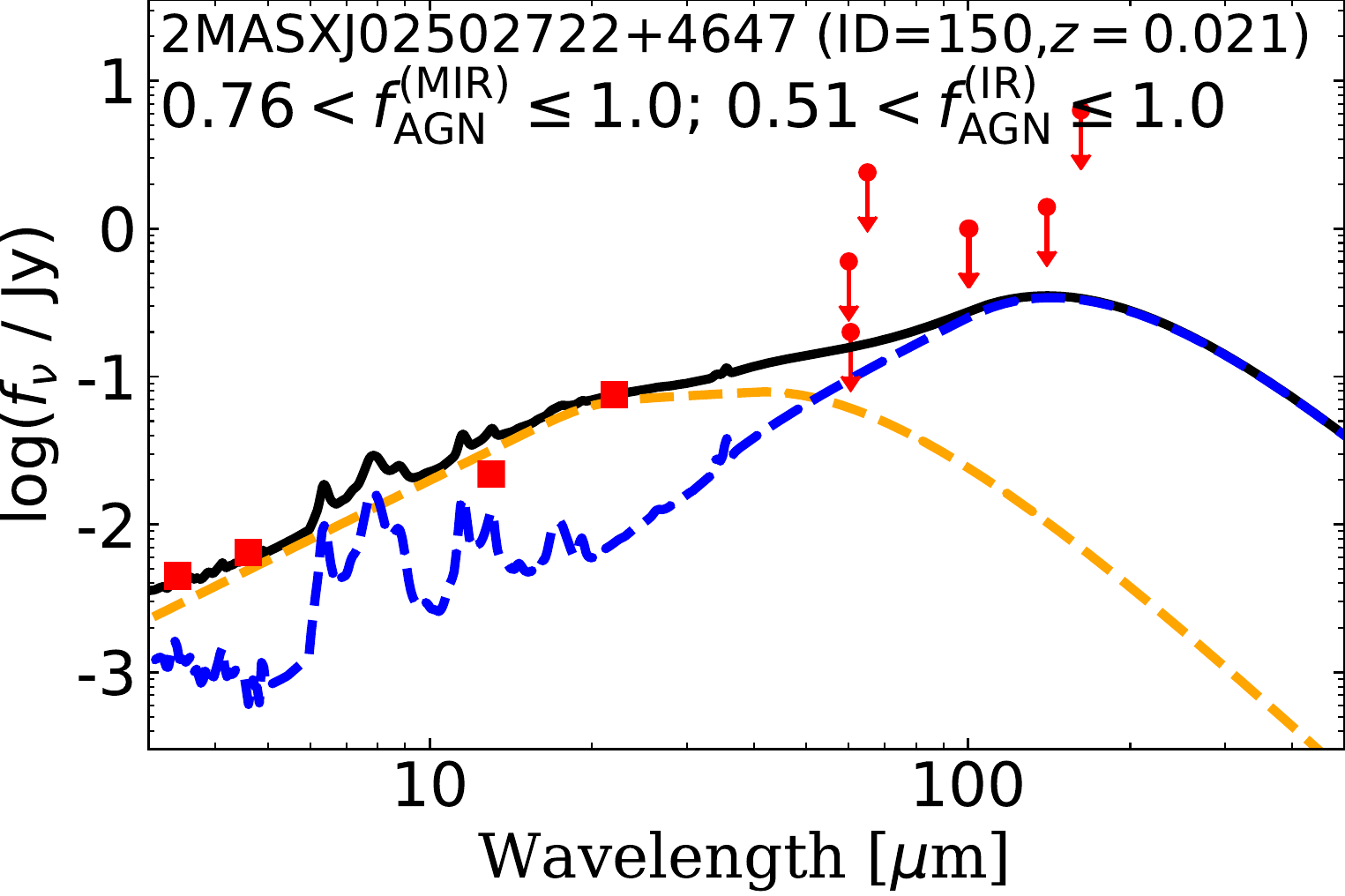}~
\includegraphics[width=0.33\linewidth]{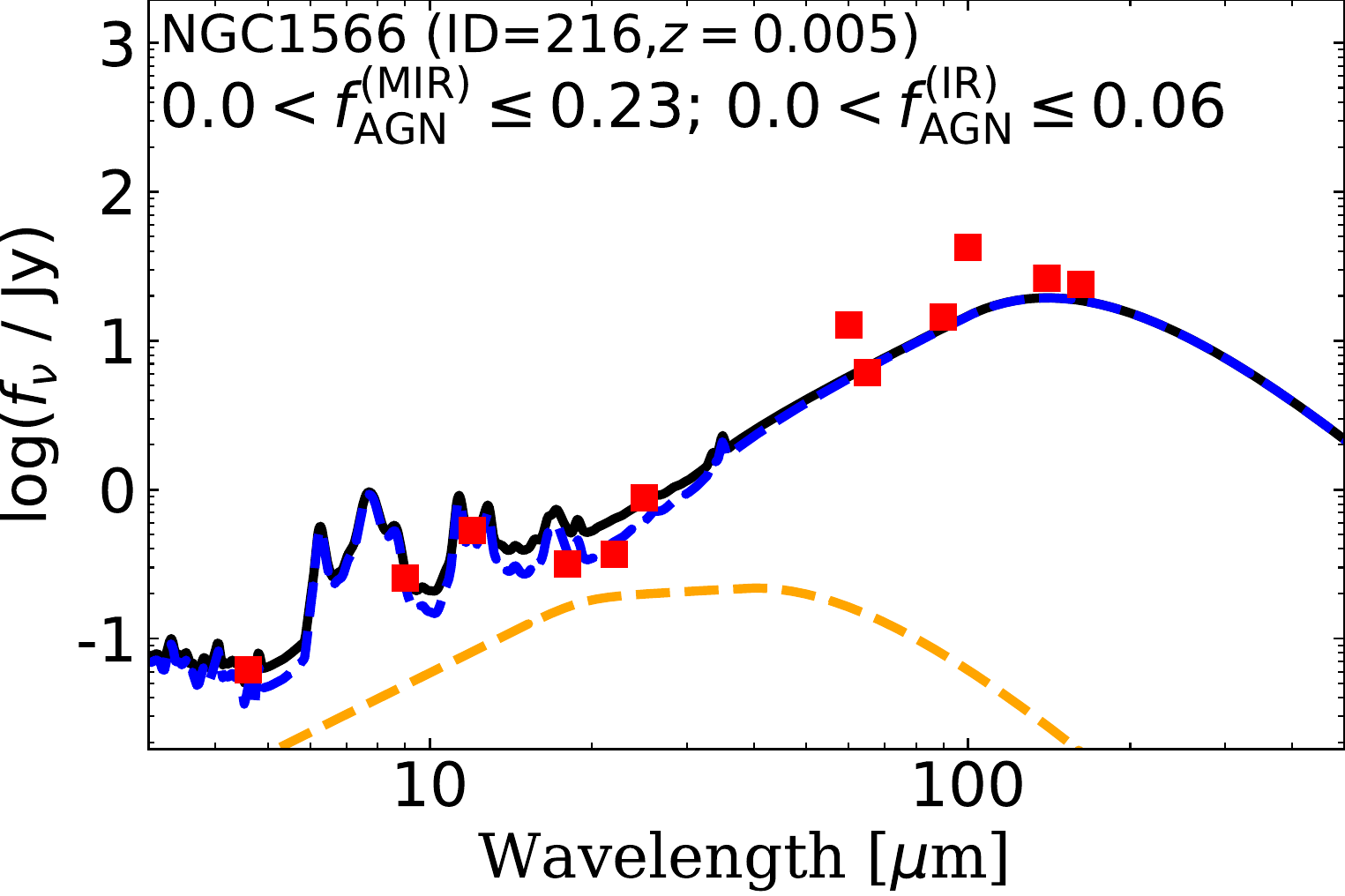}~
\caption{
Example of our IR SEDs and their best-fit models.
The orange/blue dashed curve represents fitted the AGN/host galaxy template, respectively.
The black solid curve is the combination of AGN and host galaxy template, while the red squares with error bars are the flux densities.
Each panel also shows the object ID
based on the \textit{Swift}/BAT 70-month catalog, the
redshift, and the luminosity contribution of the AGN to the MIR 
($\fagnmir$) and IR bands ($\fagnir$). 
All the SEDs of our sample are shown in the Appendix~\ref{sec:fullSED}.
Left panel: an example of a source showing both of AGN and host galaxy contributions.
Middle panel: an example of an AGN torus-dominated SED. The host galaxy template is plotted as an upper-limit.
Right panel: an example of a source with an host galaxy-dominated SED, with the AGN template plotted as an upper-limit.
}\label{fig:SED}
\end{center}
\end{figure*}

%------------------------------------------fig:SEDexample-------------------------------------%

\section{Sample}\label{sec:sample}

Our initial sample is based on the sample of \cite{ich17}, 
which contains the 606 non-blazar AGN from the \swift/BAT 
70-month catalog \citep{bau13} at galactic latitudes
($|b| > 10^{\circ}$) for which secure spectroscopic redshifts are available.
  In this study, we use the column density
and the absorption corrected 14--150~keV luminosity
 tabulated in \cite{ric16}. They are also summarized in Table~\ref{tab:IRcatalog}.
 
In \citet{ich17}, we reported the 3--500~$\mu$m IR
 counterparts for our AGN sample: 
 utilizing the IR catalogs obtained from \textit{WISE} \citep{wri10, cut13}, 
 \textit{AKARI} \citep{mur07}, \textit{IRAS} \citep{bei88}, 
and \textit{Herschel} \citep{pog10, gri10}.
Out of the 606 sources, we identified 604, 560, 601, 
and 402 counterparts in the
 total IR, near-IR ($<5~\mu$m), MIR, and far-IR
  (FIR; $60-500$~$\mu$m) band, respectively. 
    The reader should refer to \cite{ich17} for details on the IR catalogs.
  While \cite{ich17} compiled the representative fluxes at 12, 22, 70, and 90~$\mu$m,
  by combining similar wavelength bands in the multiple IR catalogs listed above, 
  in this study we regard each IR band with different central wavelength
   as independent photometry.
  Therefore, the available IR photometric bands are at most 17 bands
  between 3--500~$\mu$m,
  as identified in Table~\ref{tab:IRcatalog}.
For the data points with the same wavelengths (i.e., 12, 25, 60, 100, and 160~$\mu$m), 
the adopted photometry was chosen based on the priorities reported
in the IR catalog of \cite{ich17}
to measure the IR emission from both nucleus
and host galaxy in a uniform way for the entire AGN sample.
The 12~$\mu$m flux density was obtained with the following priority: \textit{WISE}, \textit{IRAS}/Point Source Catalog (PSC),
and \textit{IRAS}/Faint Source Catalog (FSC);
for the 25, 60, and 100~$\mu$m flux densities, on the other hand, we followed a different order (\textit{IRAS}/PSC and \textit{IRAS}/FSC), 
while for the 160~$\mu$m flux density we used
\textit{Herschel}/PACS and, when not available, \textit{AKARI}/FIS.

%%about the spatial resolution 
The corrected data are obtained from a wide range of different angular resolution
from \textit{Herschel}/PACS (70~$\mu$m; 6~arcsec) to \textit{IRAS}/FIR (100~$\mu$m; $\approx 1$~arcmin). 
Using nearly the same sample, \cite{mus14} already showed that
the bulk of PACS 70~$\mu$m is point-like at the spatial resolution of \textit{Herschel}, suggesting that the FIR emission from the host galaxy is really
compact (with a median value of 2~kpc FWHM) 
and unresolvable for most of our sample. 
Thus, we conclude that the aperture dependence with
more moderate resolutions obtained by \textit{AKARI} and 
\textit{IRAS} is negligible (see also \citealp{mel14} and \citealp{ich17}).

 %IR SEDs
 To acquire IR SEDs with a number of data points
 sufficient for spectral decomposition we require, for each source, at least 
 three photometric bands within the restframe 3--500~$\mu$m.
 This is because three data points are needed
 to define the normalization of the two components (AGN torus and host galaxy).
Applying this criteria, our sample is reduced to 588 sources.
In addition, we require at least one data point from
either the NIR or the FIR band to estimate the host galaxy component, 
which brings the sample to 587 sources.
This is the final sample used for this study, and it represents a 
large fraction of the initial sample ($587/606 = 97$\%). 
The redshift distribution of the sample is shown in Figure~\ref{fig:zdist}.\footnote
 {M81 is not shown in the figure due to its very low redshift
of $z = 10^{-4}$ \citep[see also][]{ric16}.}

%unobscured/obscured AGN
We divide the sample into two AGN types based on $N_{\rm H}$
obtained by \cite{ric16}. 
We define the AGN with $N_{\rm H} < 10^{22}$~cm$^{-2}$ as
unobscured AGN, and the AGN with 
 $N_{\rm H} \ge 10^{22}$~cm$^{-2}$ as obscured.
Overall we have 300 unobscured and 287 obscured AGN. 
The AGN type for the complete BAT 70-month catalog are tabulated in \cite{ric16},
as well as in Table~\ref{tab:IRcatalog}.
We note that \cite{kos17} found a 95\% agreement for the
unobscured and obscured AGN with the presence of a broad H$\beta$ line
for optical types Seyfert 1--1.8 and Seyfert 2.

\section{Analysis}\label{sec:SEDfitting}

We decompose the IR SEDs of AGN using SB and AGN templates
to estimate the intrinsic AGN IR luminosity.
We use the IDL script \verb|DecompIR| coded by \cite{mul11} and further developed by 
\cite{del13}.
This code accepts IR photometry points in the 3--500~$\mu$m range as input
and properly accounts for the filter and instrument response
functions of the photometry points.
It then computes
the approximate levels of AGN and host-galaxy contribution by fitting
the data combining a host-galaxy component with an AGN.
\verb|DecompIR| contains the mean AGN template produced from
the \textit{Swift}/BAT 9-month catalog \citep{tue08}, which broadly
traces the typical spectral forms of face-on and edge-on clumpy torus
models \citep[e.g.,][]{nen08a,nen08b} as shown in \cite{mul11}.
It also includes the five star-forming galaxy templates \citep{mul11, del13},
 using the average starburst SEDs derived by \cite{dal01}.
The five galaxy templates are composites of local star-forming galaxies
with $L_{\rm IR}< 10^{12}~L_{\odot}$ \citep{bra06}.
They characterize well the full range of host-galaxy SED shapes
\citep{del13,sta15}, such as the galaxy template library of \cite{cha01}.
Using these representative templates, we are able to fit the data
without suffering from the degeneracy of the fitting procedure caused
by the large number of templates.
In addition, as some of our sources have only three data points, it is reasonable to keep the number of
free parameters as small as possible.

The free parameters of the fitting are the normalizations of the AGN
and of the host-galaxy templates; therefore at least three
IR data points are needed to fit the SEDs.
However, only for the very luminous sources, we added one more free parameter.
It is known that, in high-luminosity AGN, the IR SEDs
 become much flatter at shorter wavelengths, which could be
 related to the stronger radiation field heating the surrounding dust
  to higher temperatures compared to moderate-luminosity 
  AGN \citep[e.g.,][]{ric06,net07,mul11,sym16,lyu17,lan17c}.
  Our AGN SEDs also show such tendency, especially at high luminosities
  ($L_{14-150}> 10^{44}$~erg~s$^{-1}$). 
 Therefore,  for the sources which have at least four data points and
 luminosities $L_{14-150}> 10^{44}$~erg~s$^{-1}$,
we also allow the spectral index $\alpha_{1}$ of the AGN template
\citep[see][]{mul11} to be shallower
 at wavelength shorter than 19~$\mu$m.

To determine the best fitting parameters,
we first fit the SED by using the
 five host galaxy templates (SB1--SB5) and the AGN template. 
We then check the results obtained using the five different SB templates,
and we choose the one providing the best results according to
the chi-squared statistic ($\chi^2$) minimization.

%------------------------------------------fig:AGNcontrivsLx-------------------------------------%
\begin{figure*}
\begin{center}
\includegraphics[width=0.33\linewidth]{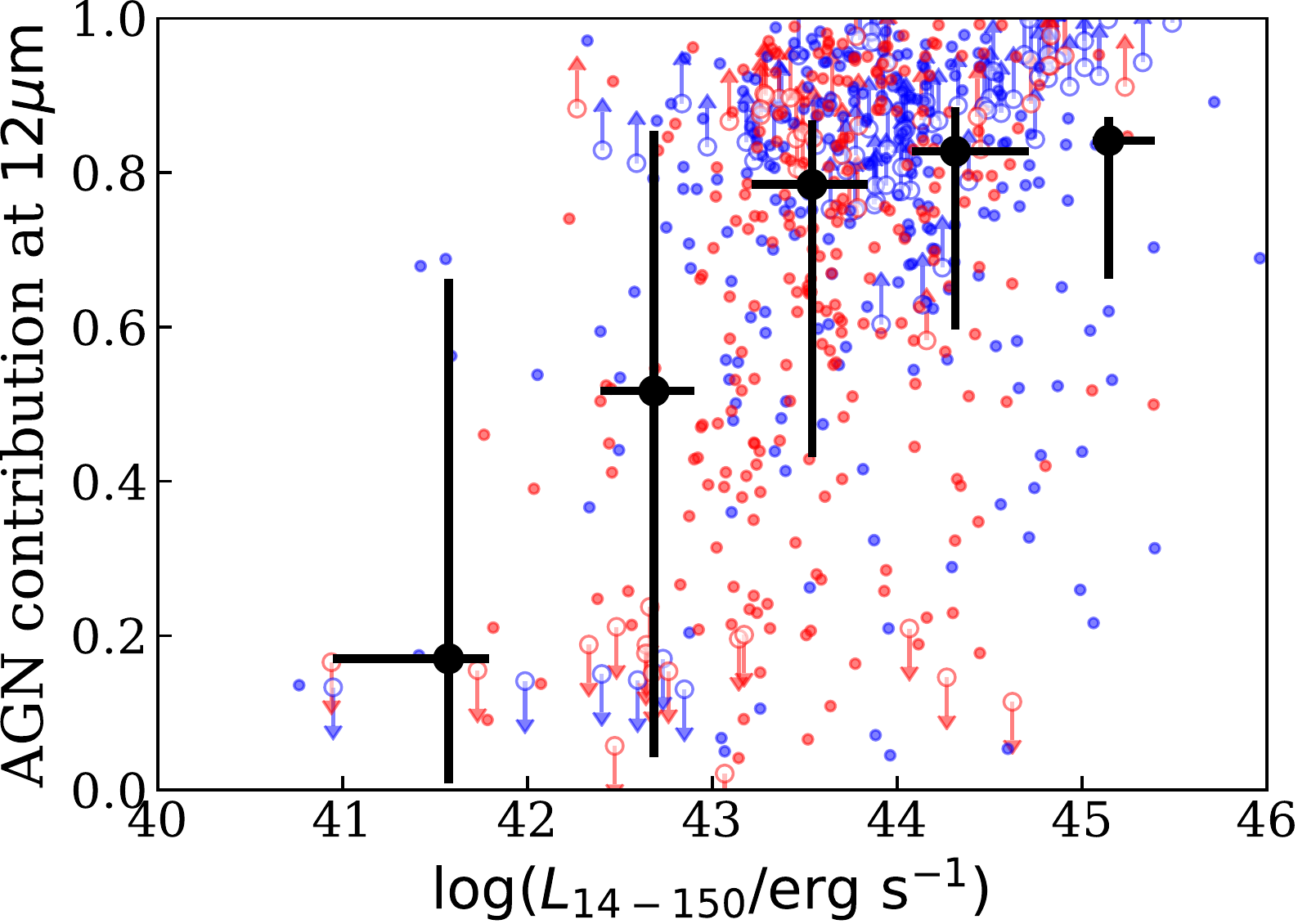}~
\includegraphics[width=0.33\linewidth]{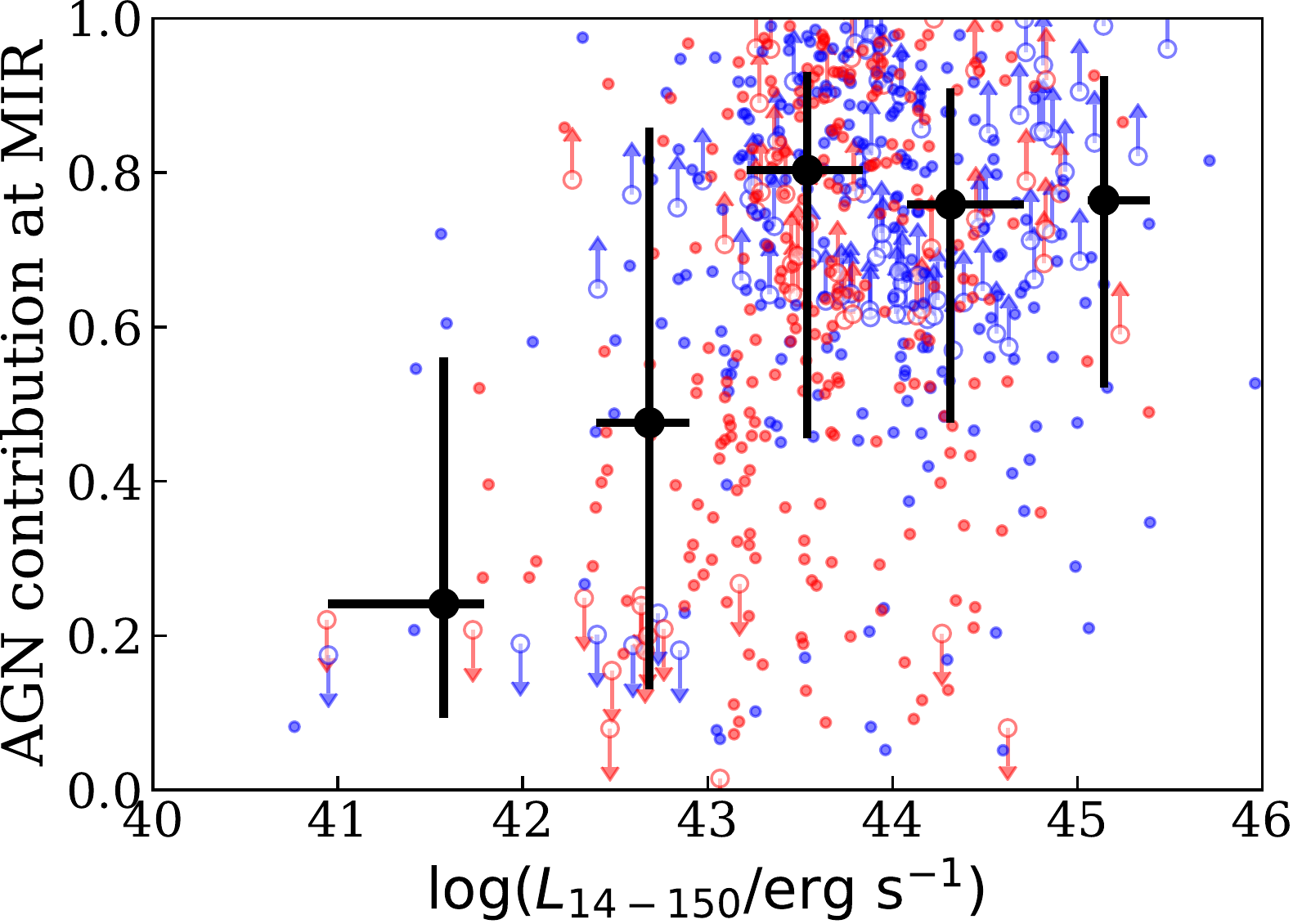}~
\includegraphics[width=0.33\linewidth]{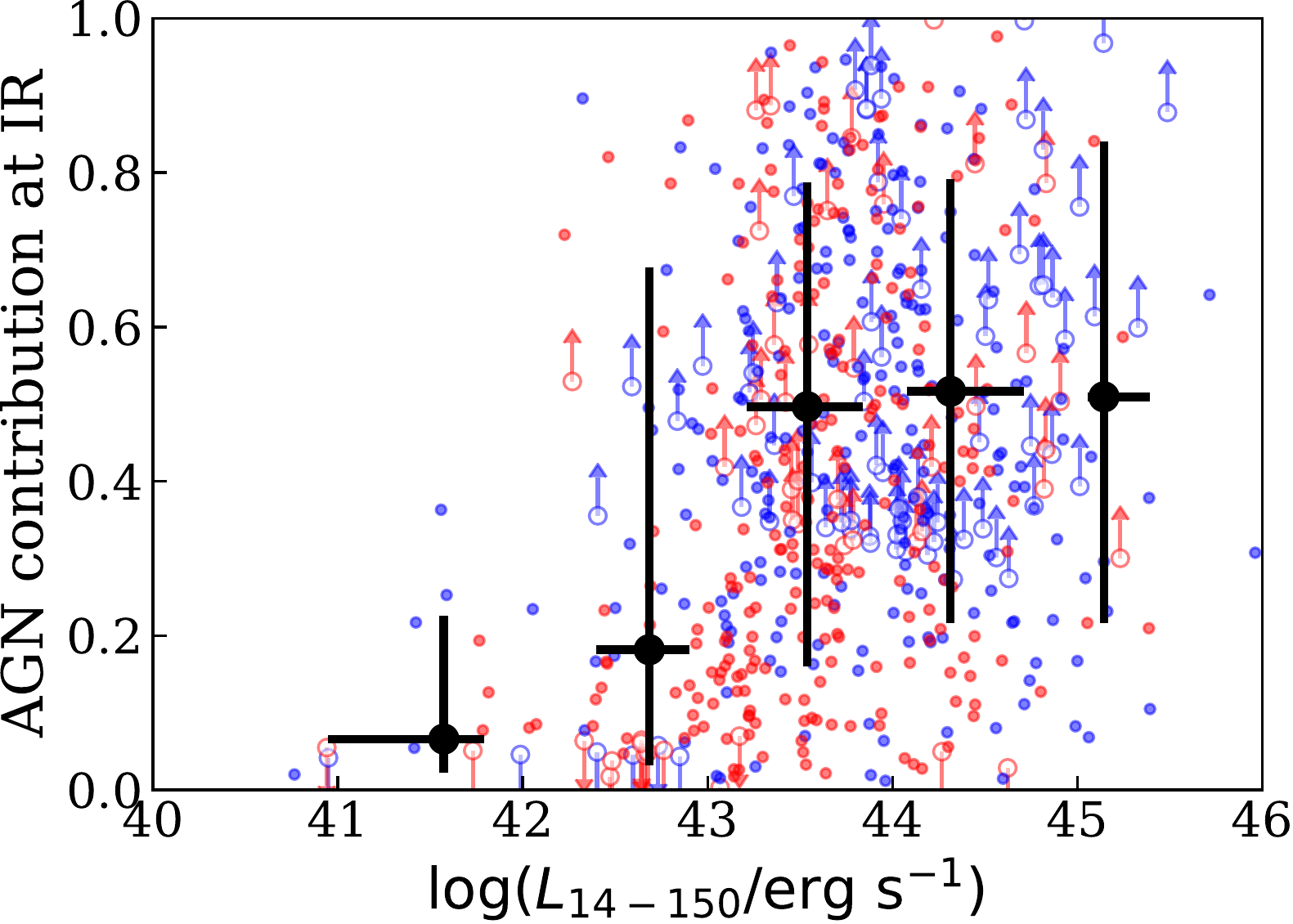}
\caption{
Fractional luminosity contribution of AGN to the
12~$\mu$m (left), MIR (middle), and total IR (right) luminosity,
as a function of 14--150~keV luminosity ($L_{14-150}$). 
The blue (unobscured AGN) and
red (obscured AGN) circles represent individual sources. 
The circles with lower/upper-limits represent the sources that require
only the AGN or host galaxy template, as discussed in Section~\ref{sec:SEDfitting}.
The black crosses represent the median contribution of the AGN luminosity 
in each bin of $L_{14-150}$, with the error bars showing the inter-percentage range with 68.2\% of the sample.
}\label{fig:AGNcontri}
\end{center}
\end{figure*}
%------------------------------------------fig:AGNcontrivsLx-------------------------------------%

%%About AGN component ==100% sources
Figure~\ref{fig:SED} shows examples of the best-fitting SEDs that
include both the AGN and star formation components, together with the best-fitting SEDs requiring only the host galaxy or the AGN component.
All the other SEDs of our sample are compiled in the Appendix~\ref{sec:fullSED}.
Overall, 474 sources required both the AGN and the host galaxy
templates, while 94 sources required only the AGN template. 
For the latter objects, the fitting quality does not improve even 
when including an additional SB template.
Since most of those sources (89 out of the 94 sources) 
are not detected in the FIR bands, and considering that the FIR bands have
 shallower sensitivities than the MIR ones,
 the lack of a significant contribution of the SB template in the MIR
  does not always imply that the host galaxy does not contribute to the total IR luminosity.
In order to assess how much the host galaxy could contribute to the total
infrared luminosity without affecting the observed SEDs, we calculate
the upper limits on the star formation contribution by following \cite{sta15},
where the same SED decomposition routine, \verb|DecompIR|, was used.
This was done by increasing the normalization of the host galaxy template until it reached
one of the upper limits, or exceeded the 3$\sigma$ uncertainty of a data point.
We then used the star-forming galaxy template giving the highest
value of IR luminosity as our conservative upper limit.
For the sources which have an upper limit on the host galaxy component, we show the
 lower limit values of the AGN contribution to the MIR
 flux ($5$--$38$~$\mu$m; $\fagnmir$) and to the total IR flux 
 (5--1000~$\mu$m; $\fagnir$)
 in each SED, as illustrated in the middle panel of Figure~\ref{fig:SED}.
The lower-limits on $\fagnmir$ and $\fagnir$ are reported in Table~\ref{tab:IRcatalog}, 
and readers can use the flag (\verb|flag_limit|) to assess whether the values are lower-limits or not.

%%SB dominated sample
There are 18 sources in our sample which were best-fit to the
host galaxy template alone ($\fagnmir=0$).
 Again, in order to assess the contribution of AGN to the total IR luminosity,
 we calculate the upper limits on the AGN torus contribution with the same methods
 of the AGN dominated SEDs, as discussed above.
 The upper limits of $\fagnmir$ and $\fagnir$
 are also shown in the right panel of Figure~\ref{fig:SED} (see also Table~\ref{tab:IRcatalog}).

Using this SED fitting approach, we have measurements of the AGN luminosity
in the 12~$\mu$m  ($\lagntwelve$), MIR ($\lagnmir$) and
 total IR bands ($L^{\rm (AGN)}_{\rm IR}$).
 All the values, as well as the IR flux densities, are tabulated in Table~\ref{tab:IRcatalog}.
We do not compile the IR star forming luminosity, 
due to the impossibility of obtaining a reliable estimates for the sources not detected in the FIR.

\section{Results and Discussion}

\subsection{Fractional Luminosity Contribution of AGN to the IR Band}

Figure~\ref{fig:AGNcontri} shows the median of the AGN contribution
to the 12~$\mu$m,  MIR, and total IR luminosities as 
a function of $L_{14-150}$.
The AGN contribution is calculated from the ratio between the AGN and the total (SF plus AGN) luminosity:
\begin{equation}
 f_{\rm AGN}^{(12 \mu{\rm m, MIR, IR})} = 
 L_{12 \mu{\rm m, MIR, IR}}^{(\rm AGN)}/ (L_{12 \mu{\rm m, MIR, IR}}^{(\rm AGN)} + L_{12 \mu{\rm m, MIR, IR}}^{(\rm SF)}).
\end{equation}

Figure~\ref{fig:AGNcontri} shows that the luminosity contribution of 
the AGN to the 12~$\mu$m, MIR, and to the total IR band increases with $L_{14-195}$.
On the low-luminosity end ($L_{14-195}<10^{43}$~erg~s$^{-1}$), Figure~\ref{fig:AGNcontri}
 indicates that the host galaxy emission significantly contaminates 
 ($\simeq 50$--$80$\%)
 the 12~$\mu$m and MIR band.
On the high luminosity end ($L_{14-195}> 10^{43}$~erg~s$^{-1}$), it clearly
shows that the AGN component is the dominant ($\gtrsim80$\%) energy source at 12~$\mu$m and in the MIR band. 
This overall result is broadly consistent with the 
previous studies which explored the AGN contribution 
using high spatial resolution imaging
\citep[e.g.,][and references therein]{asm11,asm14}.
These works are discussed in the Appendix~\ref{sec:CompAsm14}.
 On the other hand, in the total IR band,
 the AGN component contributes only up to $\simeq 50$\% even
  at high luminosities. This result is consistent with the calculations of local quasars \citep{lyu17},
  where it is shown that AGN contribute to $\simeq 50$\% of the flux even if they
  provide 90\% of the MIR emission.

Figure~\ref{fig:AGNcontri} also shows that
the scatter of the percentage for $\fagnmir$
is $\sim 20$\% and increases up to $\sim 35$\% for $\fagnir$. 
The origin of the scatter is mostly due to
AGN-dominated sources without any detections in the FIR bands.
 Since the distant sources with $z>0.05$ have not been observed 
 with \textit{Herschel} \citep[see][]{mel14, shi16},
those sources have very shallow upper limits: 0.2~Jy at 60~$\mu$m (\iras/FSC)
and/or 0.55~Jy at 90~$\mu$m (\akari/FIS). This allows a possible contribution of the host galaxy emission to the FIR bands,
  even when its contribution to the MIR flux is negligible
  as discussed in Section~\ref{sec:SEDfitting} \citep[see also ][]{lyu17}.
Therefore, higher sensitivity FIR photometry is crucial to quantify
 the host galaxy contribution for those sources.
 
  %------------------------------------------fig:L2color-------------------------------------%
\begin{figure}
\begin{center}
\includegraphics[width=9.0cm]{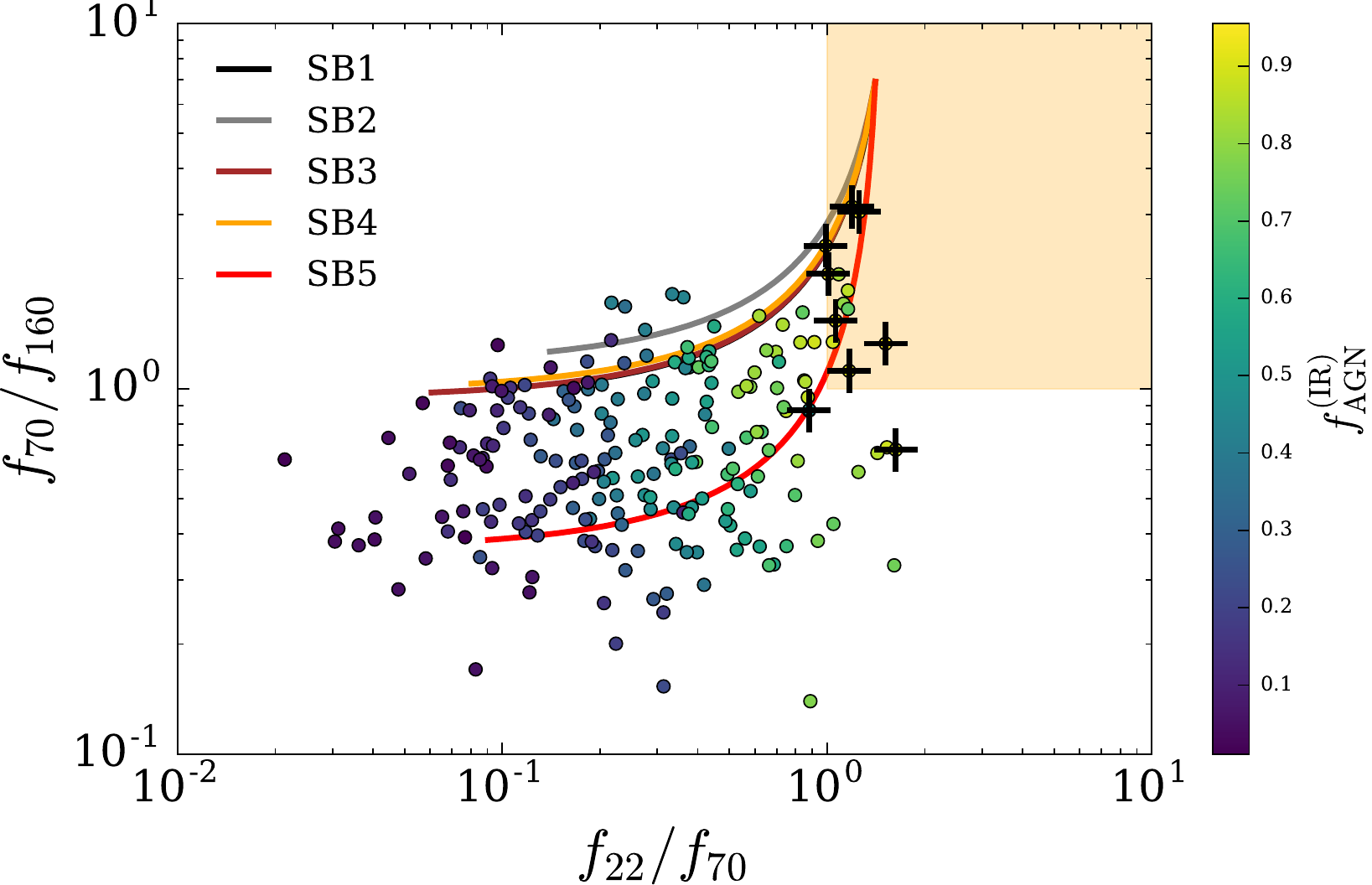}~
\caption{
Observed 
$f_{70~\mu{\rm m}}/f_{160~\mu{\rm m}}$ versus 
$f_{22~\mu{\rm m}}/f_{70~\mu{\rm m}}$
ratio for sample sources with secure detections 
in the $22$~$
 \mu$m, $70$~$\mu$m, and $160$~$\mu$m bands.
 The color-color variations as a function of $\fagnir$ 
 are also plotted for the five SB templates used in this study
 originated from \cite{mul11}.
 The black crosses are the IR pure-AGN sources discussed 
 in Section~\ref{sec:pureAGN}.
 The colorbar represents the AGN contribution to the total IR band ($\fagnir$).
 The orange area illustrates the region with $f_{22~\mu{\rm m}}>f_{70~\mu{\rm m}}>f_{160~\mu{\rm m}}$.
 }\label{fig:L2color}
\end{center}
\end{figure}
%------------------------------------------fig:L2color-------------------------------------%

%------------------------------------------fig:SEDexample-------------------------------------%
\begin{figure*}
\begin{center}
\includegraphics[width=0.33\linewidth]{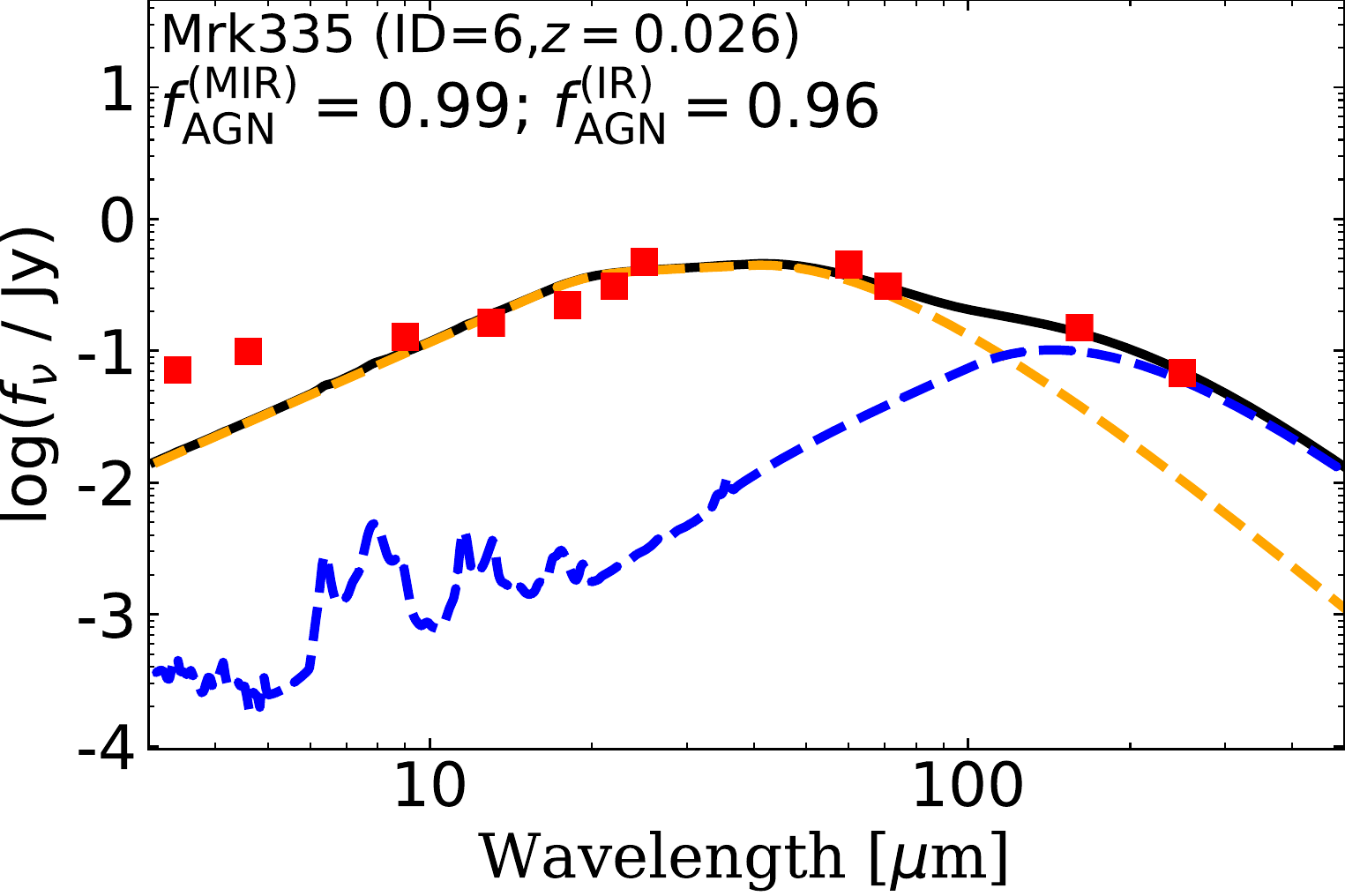}~
\includegraphics[width=0.33\linewidth]{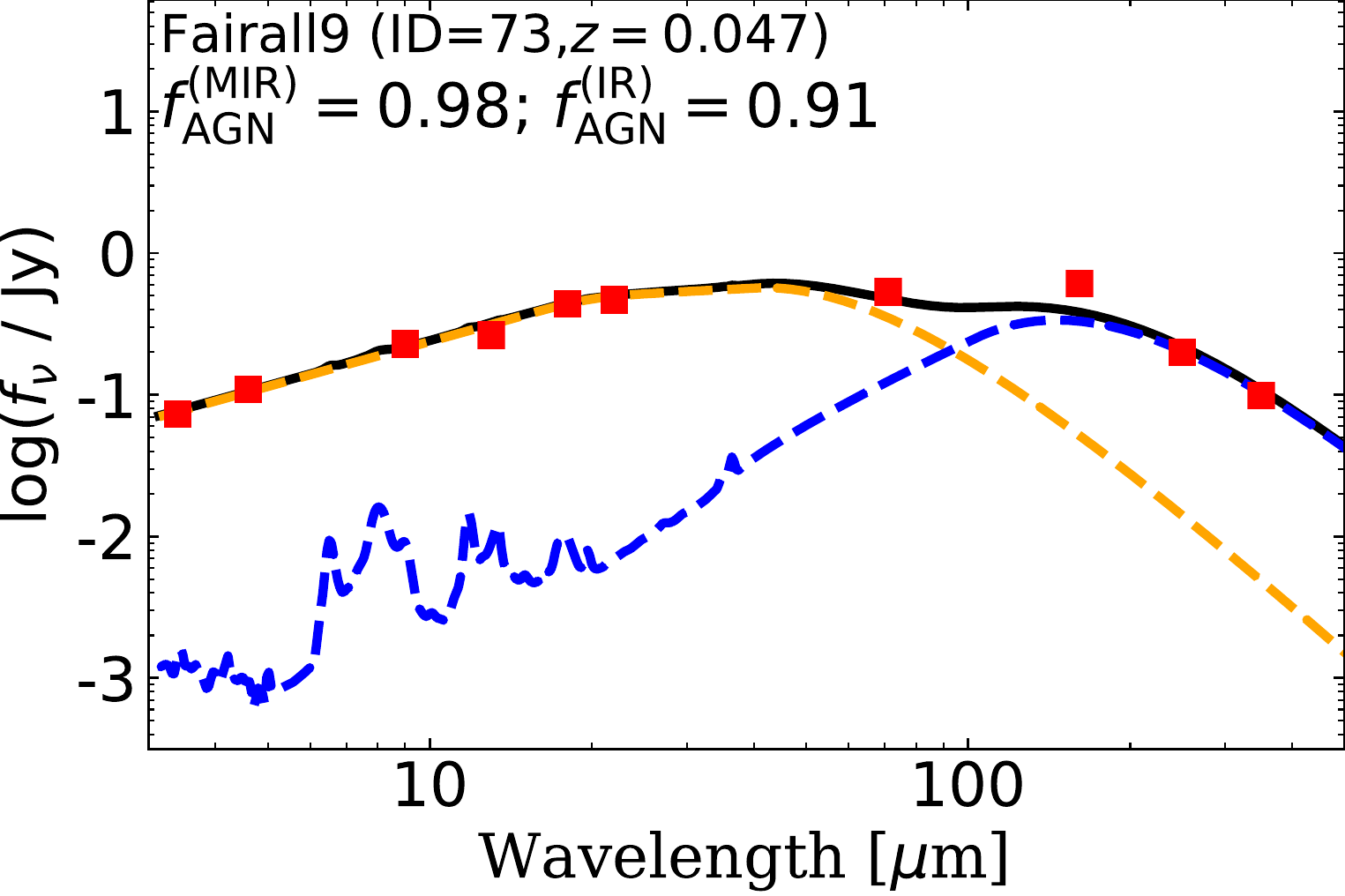}~
\includegraphics[width=0.33\linewidth]{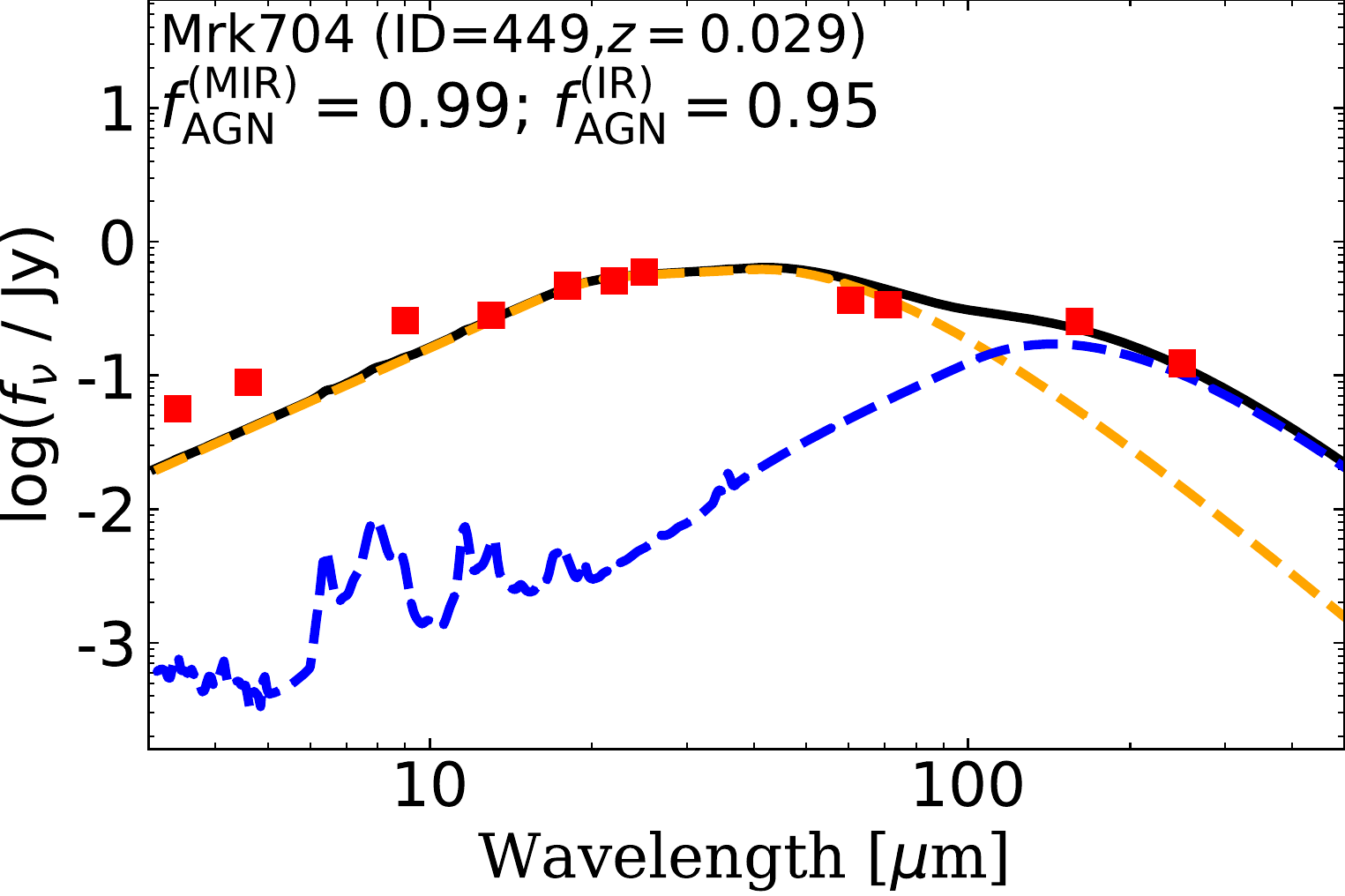}\\
\includegraphics[width=0.33\linewidth]{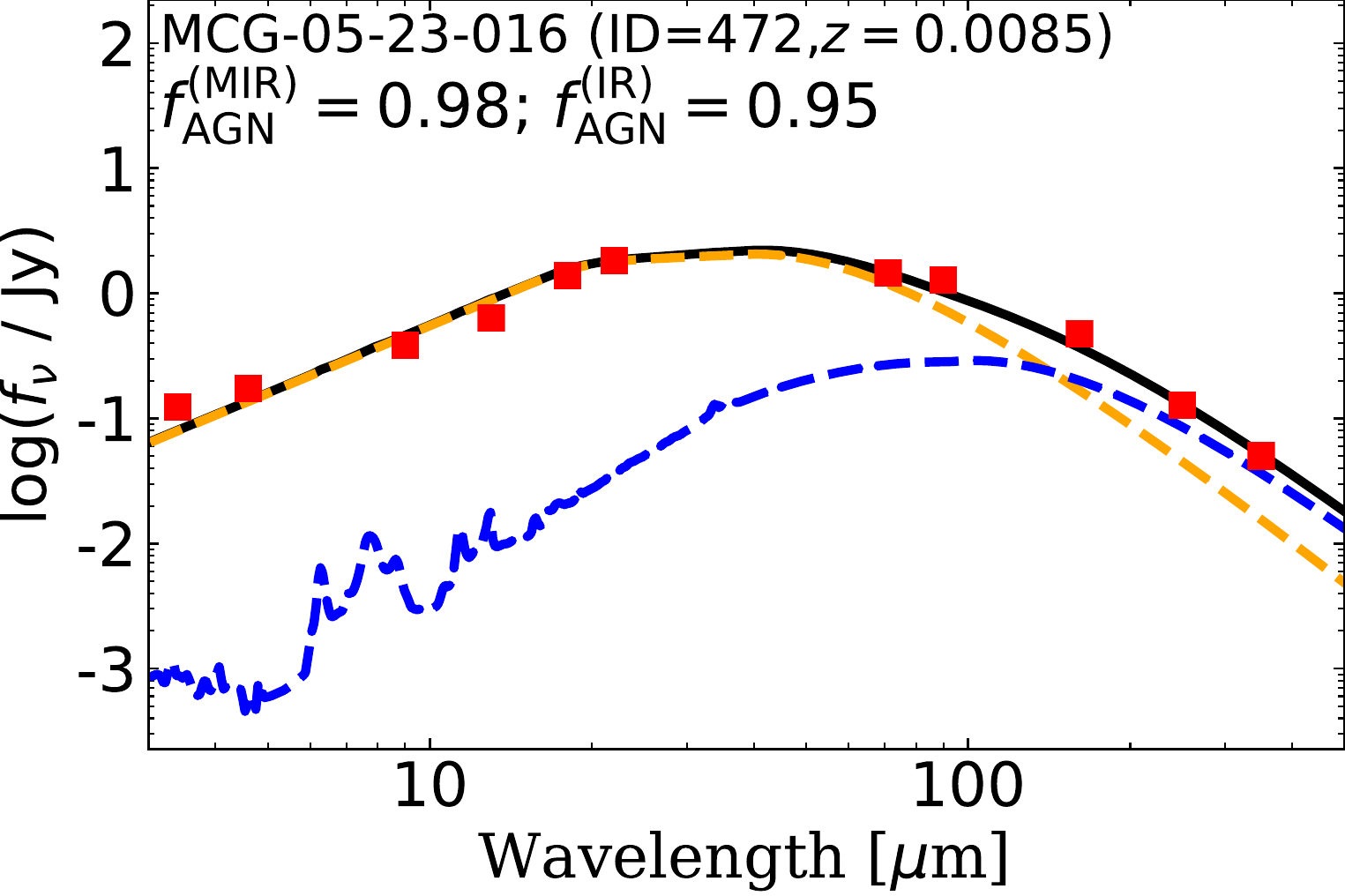}~
\includegraphics[width=0.33\linewidth]{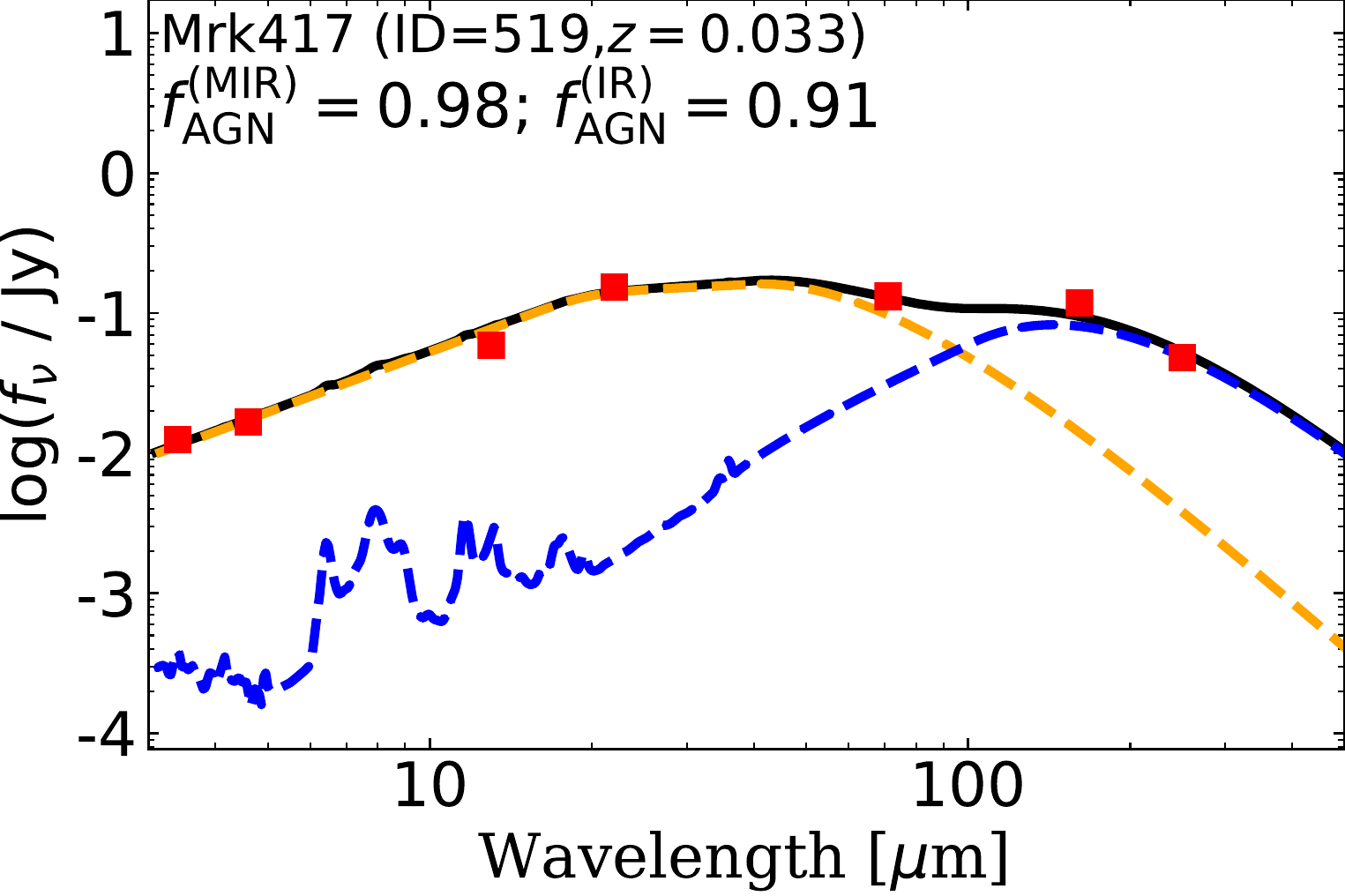}~
\includegraphics[width=0.33\linewidth]{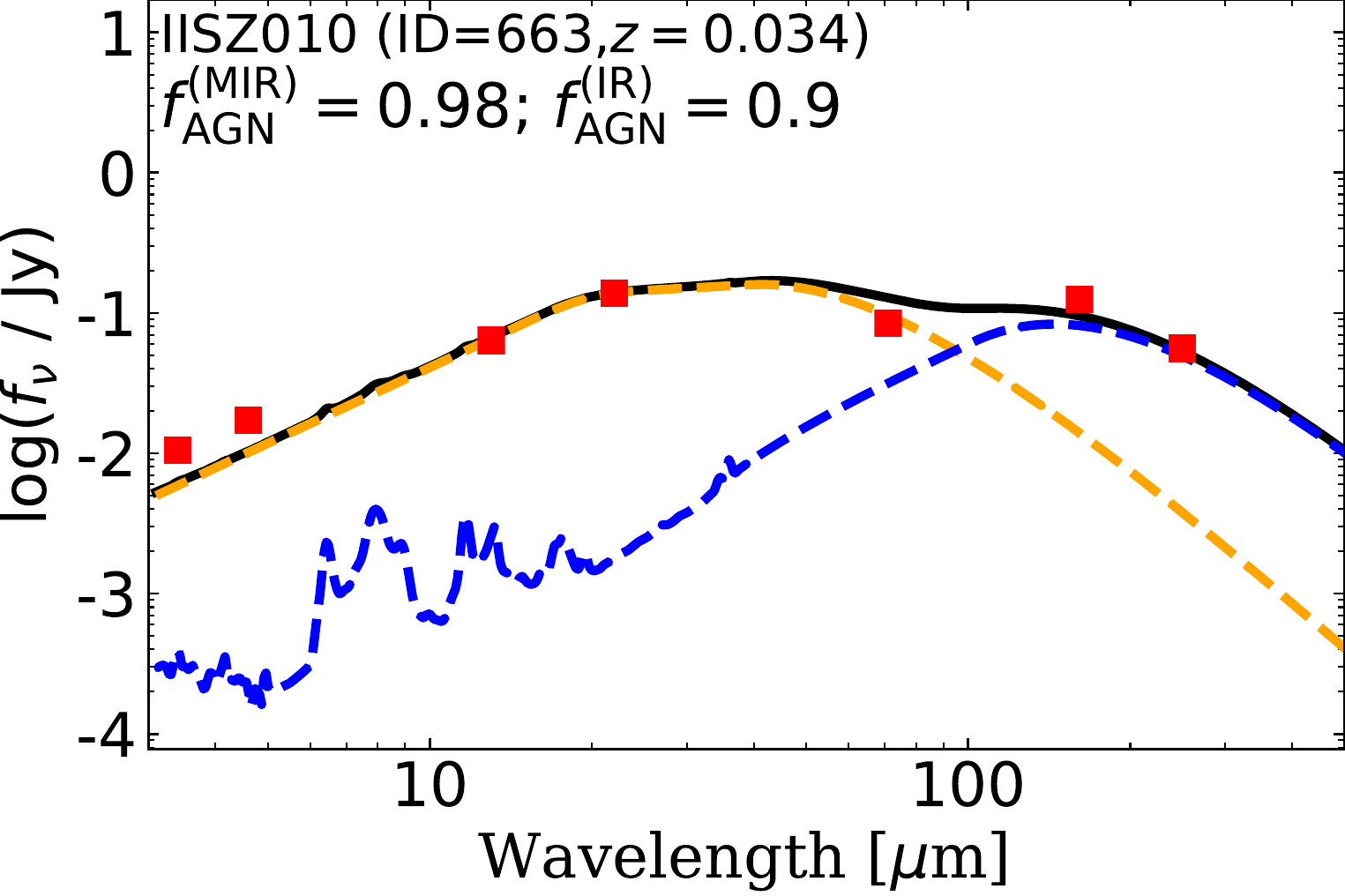}\\
\includegraphics[width=0.33\linewidth]{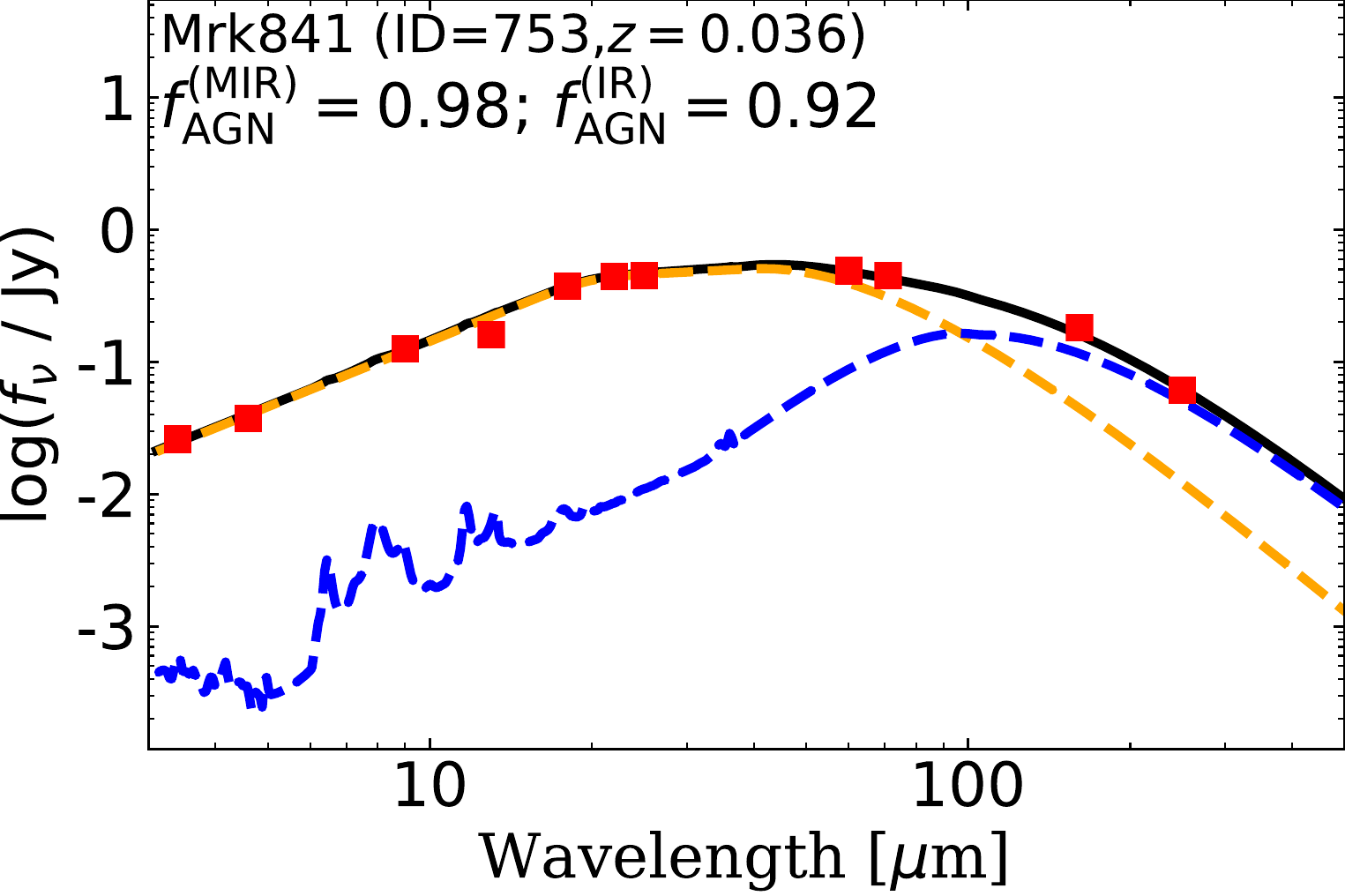}~
\includegraphics[width=0.33\linewidth]{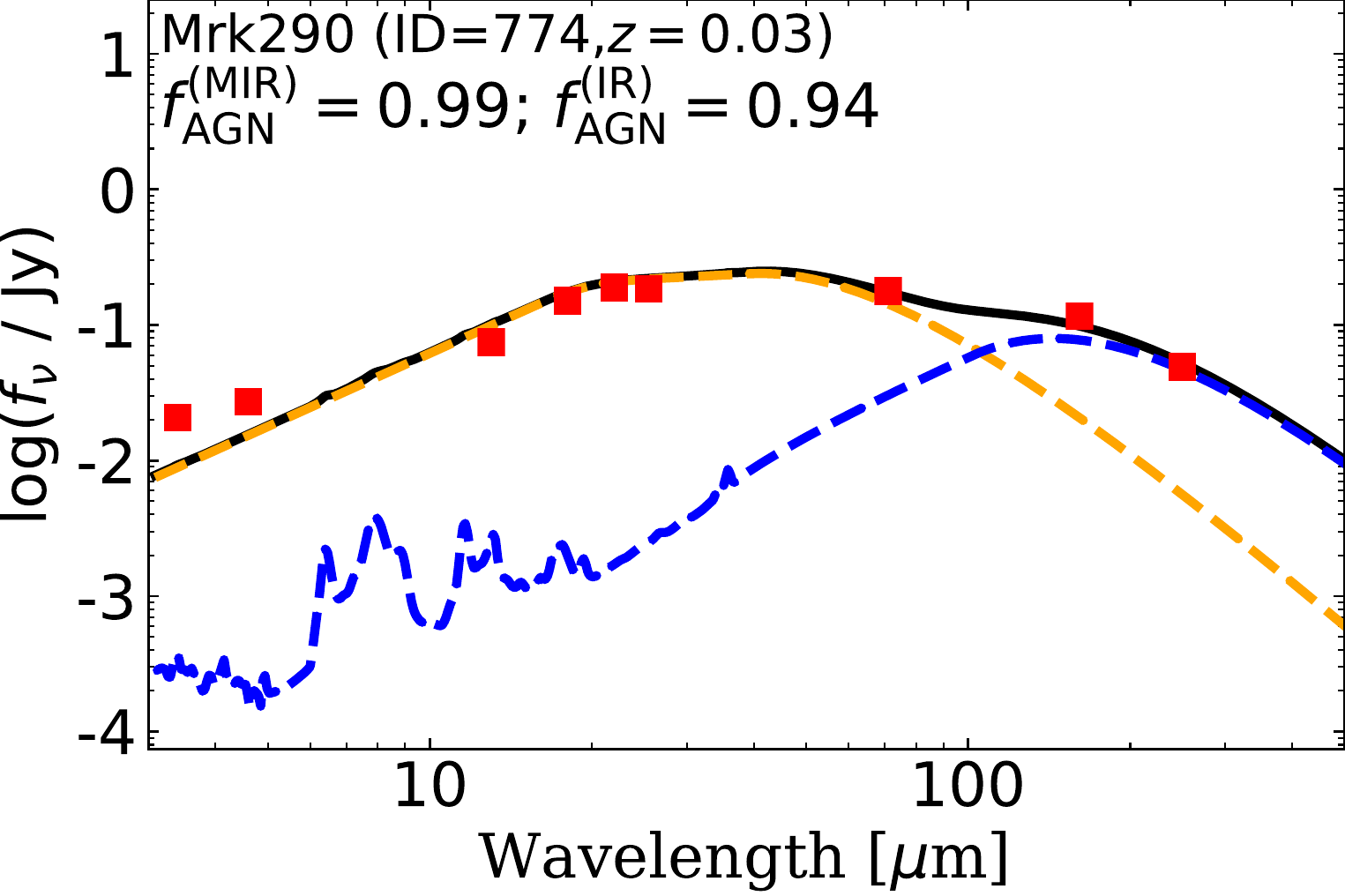}~
\includegraphics[width=0.33\linewidth]{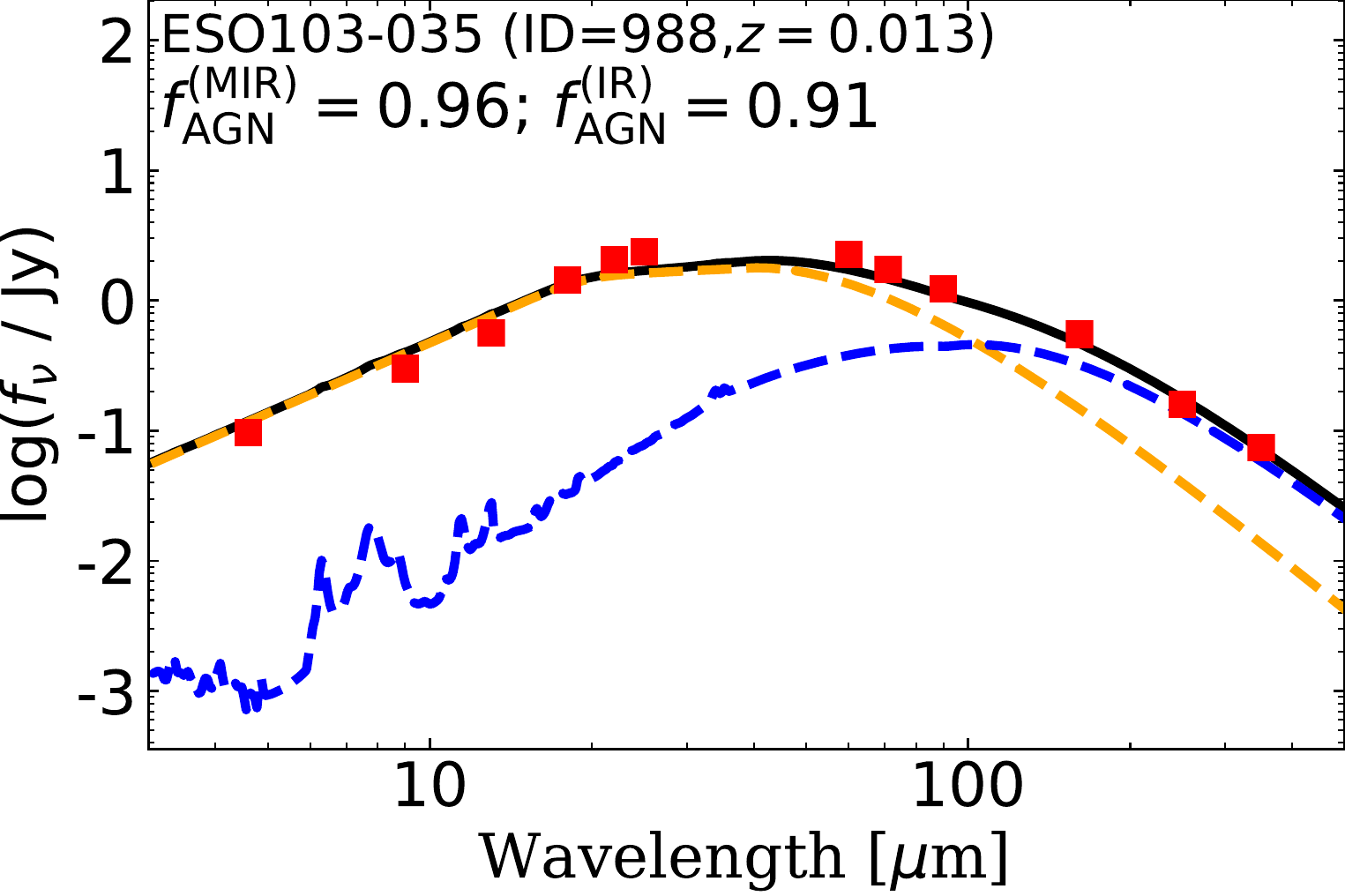}~
\caption{
SEDs of the IR pure-AGN candidates defined as
1) $f_{\rm AGN}^{(\rm IR)}>0.90$ and 2) significant detection both at 60--70~$\mu$m and 160~$\mu$m.
 All plots are same as in Figure~\ref{fig:SED}.
}\label{fig:pureAGNSED}
\end{center}
\end{figure*}
%------------------------------------------fig:SEDexample-------------------------------------%

\subsection{IR Pure-AGN Candidates}\label{sec:pureAGN}

%Introduction of what pure-IR AGN is:
Some sources show AGN-dominated SEDs even in the FIR bands.
These sources are called IR pure-AGN \citep{mul11,ros12, mat15a, ich17,ros18},
and are ideal candidates to derive intrinsic AGN IR templates.
%Introduction of SED feature of pure-IR AGN
These IR pure-AGN have a spectral turn-over 
 at 20--40~$\mu$m \citep{alo12,hon14,ful16,lop18}, and
a declining flux density from 40~$\mu$m to 160~$\mu$m,
suggesting a very low contribution from the starburst in the host galaxy.
%Introduction of color-color diagram.
In order to check the SED turn-over quantitatively, 
we plot IR color-color plots of 
$f_{70~\mu{\rm m}}/f_{160~\mu{\rm m}}$ versus $f_{22~\mu{\rm m}}/f_{70~\mu{\rm m}}$ in Figure~\ref{fig:L2color}.
Both flux ratios are known to be sensitive to the SED peak, and therefore to the dust temperature \citep{mel14,gar16}. 
 The orange shaded area in Figure~\ref{fig:L2color} 
 ($f_{70~\mu{\rm m}}/f_{160~\mu{\rm m}}>1.0$ 
and $f_{22~\mu{\rm m}}/f_{70~\mu{\rm m}}>1.0$) indicates a decline in flux density as a function of wavelength from 22~$\mu$m
to 160~$\mu$m since the sources fulfill $f_{22~\mu{\rm m}}>f_{70~\mu{\rm m}}>f_{160~\mu{\rm m}}$.

Figure~\ref{fig:L2color} also shows the simulated IR color as a function of 
$\fagnir$ for the each SB template.
All IR colors follow a similar trend;
$f_{22~\mu{\rm m}}/f_{70~\mu{\rm m}}$ 
increases up to $f_{22~\mu{\rm m}}/f_{70~\mu{\rm m}} \simeq 1.0$ with 
$\fagnir$ up to 0.9, while $f_{70~\mu{\rm m}}/f_{160~\mu{\rm m}}$
shows a very shallow increase until $\fagnir \leq 0.8$. However, for $\fagnir>0.9$,
 $f_{70~\mu{\rm m}}/f_{160~\mu{\rm m}}$
starts to drastically increase, reaching values up to $\simeq 7.0$.
 Thus, sources located in the orange shaded area should have 
 AGN-dominated IR SEDs with $f_{\rm AGN}^{(\rm IR)}>0.90$.
 In this study we define IR pure-AGN when the source fulfills the following criteria:
 1) $f_{\rm AGN}^{(\rm IR)}>0.90$ and
 2) a significant detection both at 60--70~$\mu$m and 160~$\mu$m.
A total of nine sources are selected with these criteria, and are shown
with the black crosses in Figure~\ref{fig:L2color}.
Most IR pure-AGN are successfully
located in the orange shaded area in the color-color-plot.
Figure~\ref{fig:pureAGNSED} shows the SEDs of the selected 9 IR pure-AGN.
All sources show an SED turn-over between $\simeq 20$~$\mu$m and $\simeq 70$~$\mu$m, a declining flux density from 70~$\mu$m to 160~$\mu$m, and do not show any FIR bump due to star formation up to 90~$\mu$m, 
with the exception of Fairall~9 and II SZ 010.
Some of the sources of our sample have already been 
reported as being dominated by torus emission in the IR from the study of their \textit{Spitzer}/IRS spectra \citep[e.g., MCG -05-23-16;][]{ich15}, based on the spectral turn-over at 20--40~$\mu$m \citep{alo12,hon14,ful16,lop18}.

%Properties of IR pure-AGN
We also check the AGN properties of IR pure-AGN compared to the parent sample.
The means and standard deviations of the logarithmic X-ray luminosity, black hole mass, and Eddington ratio
of this subsample are
$\langle \log L_{14-150} \rangle = 43.7 \pm 0.3$
$\langle \log M_{\rm BH} \rangle = 7.8 \pm 0.5$, 
and 
$\langle \log \lambda_{\rm Edd} \rangle=-1.2 \pm 0.3$, respectively.
These values are consistent with those of the parent sample
of $\langle \log L_{14-150} \rangle=43.7 \pm 0.8$, 
$\langle \log M_{\rm BH} \rangle=8.0 \pm 0.8$, 
and $\langle \log \lambda_{\rm Edd} \rangle=-1.5 \pm 0.8$.
This result suggests that the dominating AGN contribution to the total IR band
is not related to their higher AGN luminosities, lower BH masses, or higher Eddington ratio, while it implies that they have weaker star-formation luminosities
than other AGN with similar luminosities.
Actually, MCG -05-23-16 is one of the pure IR-AGN whose CO emission
has not been detected \citep{ros18} in the \textit{Swift}/BAT AGN
subset of the LLAMA survey \citep{dav15}.
This suggests that its host galaxy already lacks the molecular gas
to produce the star formation.
The on-going molecular gas observations conducted by the BASS survey
(M. Koss et al. in prep.) will explore the origin
of the deficit of star formation in these IR pure-AGN sources.

  %------------------------------------------Figure~LIRvsLx-------------------------------------%
\begin{figure*}
\begin{center}
\includegraphics[width=0.5\linewidth]{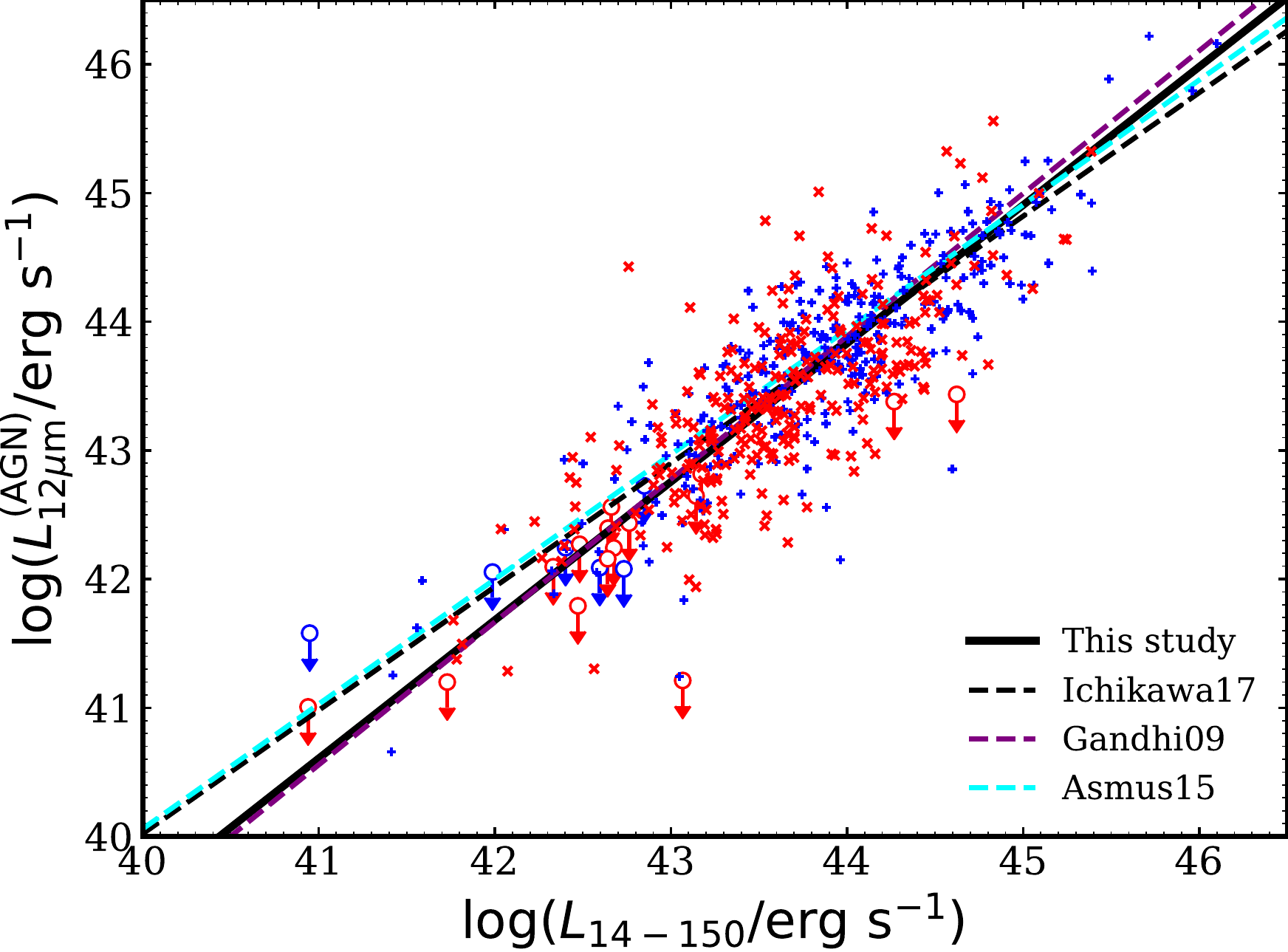}~
\includegraphics[width=0.5\linewidth]{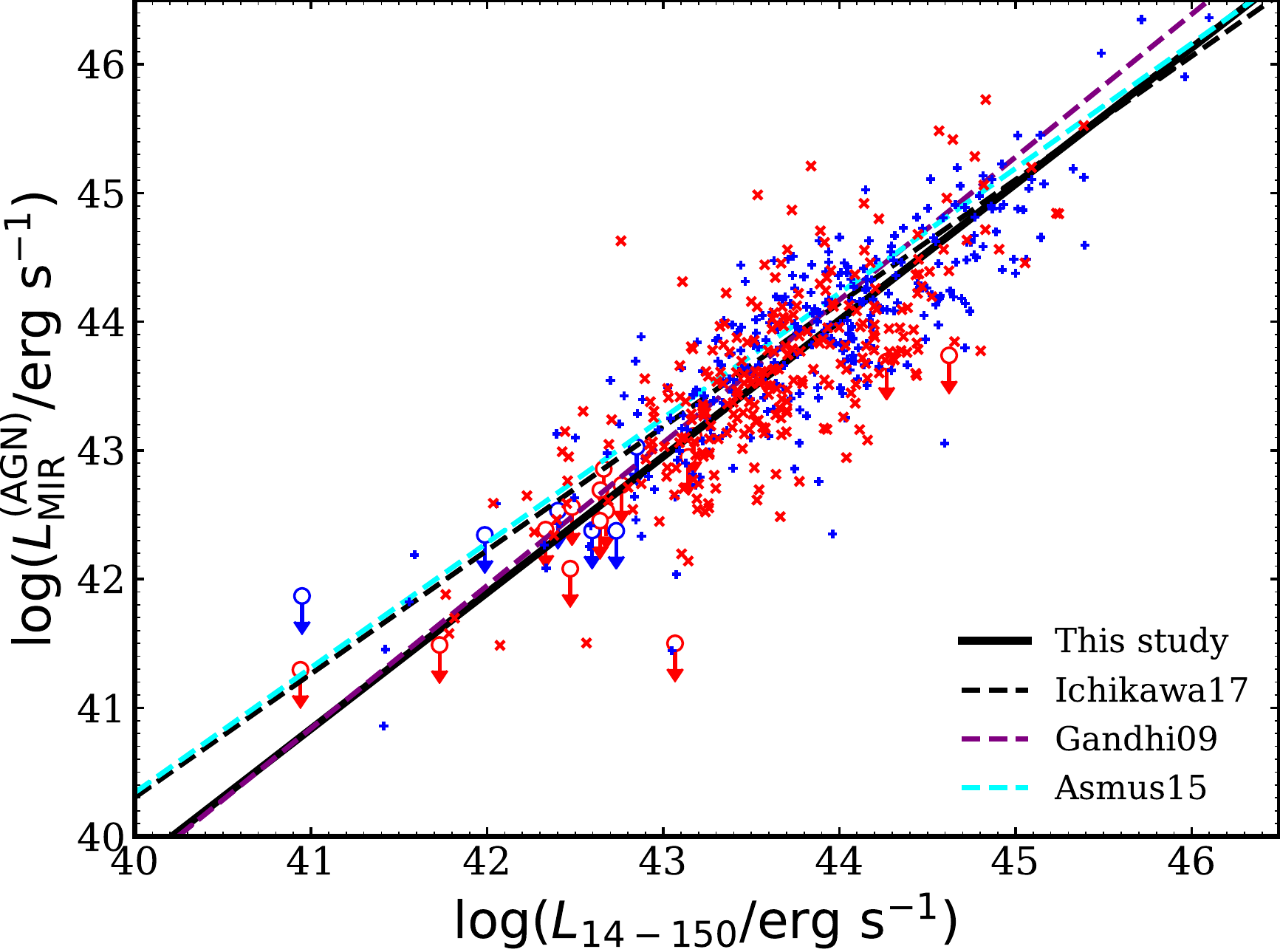}~
\caption{
Scatter plot of the 14--150~keV ($L_{14-150}$) luminosity and the AGN 12~$\mu$m ($L^{\rm (AGN)}_{12~\mu{\rm m}}$; left panel) and MIR ($L^{\rm (AGN)}_{\rm MIR}$; right panel) luminosities .
Blue and red cross represents unobscured and obscured AGN, respectively.
The black solid line represents the slope obtained by this study using the infrared bands after the SED decomposition.
The other slope represents the one obtained by our previous study;
 before the SED decomposition
\citep[][; black dashed line]{ich17}, and the higher spatial resolution studies of \cite{gan09} (purple), and \cite{asm15} (cyan).
}\label{fig:LIRvsLx}
\end{center}
\end{figure*}
%------------------------------------------Figure~LxvsLIR-------------------------------------%

\subsection{Correlation Between the 12~$\mu$m AGN and 14--150~keV Luminosities}\label{sect:LxvsLIR}

Figure~\ref{fig:LIRvsLx} shows the relation between 
$L^{\rm (AGN)}_{12 \mu {\rm m}}$, $L^{\rm (AGN)}_{\rm MIR}$, and $L_{14-150}$ in the 
$10^{40} < L_{14-150}<10^{47}$~erg~s$^{-1}$ range.
Blue and red crosses represent unobscured and obscured AGN, respectively.
The upper limits, shown as open circles, represent the 
host galaxy dominated sources which have a possible 
AGN contribution in the 12~$\mu$m and MIR band as discussed in 
Section~\ref{sec:SEDfitting}.
 
 %Slope using the ASURV
 The slope of the relation between
 $L^{\rm (AGN)}_{12 \mu{\rm m}}$, $L^{\rm (AGN)}_{\rm MIR}$ and $L_{14-150}$ is 
 estimated considering the two variables as independent parameters.
 Since our data contains both detections and upper limits,
we apply the survival analysis method using the Python 
package\footnote{\url{http://python-asurv.sourceforge.net/}} of ASURV
\citep{fei85, iso86, lav92}
to account for the upper limits on  $L^{\rm (AGN)}_{12 \mu{\rm m}}$
 and $L^{\rm (AGN)}_{\rm MIR}$. 
 We use the slope Bisector fits, which minimize perpendicular 
distance from the slope line to data points.
The fits, with the form of
 $[ \log (L^{(\rm AGN)}_{12 \mu {\rm m, MIR}}/10^{43}~{\rm erg}~{\rm s}^{-1}) =
  (a \pm \Delta a) + (b \pm \Delta b) \log (L_{14-150}/10^{43}~{\rm erg}~{\rm s}^{-1}$)],
where $\Delta a$ and $\Delta b$ are the standard deviations of $a$ and $b$, respectively,
result in
\begin{align}
\log \frac{L^{\rm (AGN)}_{12 \mu{\rm m}}}{10^{43}~{\rm erg}~{\rm s}^{-1}} &= (-0.24 \pm 0.03) + (1.08 \pm 0.03)\notag \\
& \times \log \frac{L_{\rm 14-150}}{10^{43}~{\rm erg}~{\rm s}^{-1}},\\
\log \frac{L^{\rm (AGN)}_{\rm MIR}}{10^{43}~{\rm erg}~{\rm s}^{-1}} &= (-0.05 \pm 0.03) + (1.06 \pm 0.03)\notag \\
& \times \log \frac{L_{\rm 14-150}}{10^{43}~{\rm erg}~{\rm s}^{-1}},
\end{align}
and they are also summarized in Table~\ref{tab:Equation}.
We find that both luminosity-luminosity and flux-flux correlations
are significant (see also Appendix~\ref{sec:fIRvsfbat}
for the flux-flux correlations).

In Figure~\ref{fig:LIRvsLx}, some of the fits reported by recent works are also overplotted.
Since most previous studies used the 2--10 keV luminosity, 
we apply a conversion factor of $L_{14-150}/L_{2-10}=2.36$ under the 
assumption of  the photon index $\Gamma = 1.8$, which is the median value of the 
\textit{Swift}/BAT 70-month AGN sample
\citep{ric16}, for overplotting in the same Figure. 
Since the AGN template used in this study has a ratio of
$\lagnmir/\lagntwelve = 1.92$, we also apply it to the slopes
from the previous studies
for overplotting in the relation between $\lagnmir$ and $L_{14-150}$.

Compared to \cite{ich17}, where we found $b=0.96 \pm 0.02$,
the sample used here shows a smaller 12~$\mu$m contribution from AGN
in the low-luminosity end. 
This is because the sources with lower $L_{14-195}$ have a significant host galaxy
 contamination even in the MIR, as shown in Figure~\ref{fig:AGNcontri}
 and also in the right panel of Figure~\ref{fig:A14vsDecomp}.
 Indeed, \cite{ich17} also reported that the slope becomes slightly steeper with $b=1.05 \pm 0.03$
 when one considers sources with $L_{14-195}>10^{43}$~erg~s$^{-1}$,
 for which the host galaxy contamination in the MIR is negligible.
 This is also consistent with the value of $b=1.08\pm0.03$ in this study.

We compare our results with what was found by
\cite{gan09} and \cite{asm15} using
 high spatial resolution observations of X-ray selected AGN down to the low-luminosity end.
  The MIR emission in those studies is most likely dominated from the AGN torus 
  and have a relatively low level of the host galaxy contamination 
  thanks to their spatially resolved images.
As shown in Figure~\ref{fig:LIRvsLx}, our study finds a similar slope to that reported in \cite{gan09} ($b=1.11 \pm 0.07$), and also
 within $3\sigma$ uncertainty to 
 that of \cite{asm15} ($b=0.97 \pm 0.03$).
 This strongly supports that our SED decomposition method
 nicely reproduces the high spatial resolution flux, which 
 is thought to  be dominated from AGN torus emission.

%------------------------------------------Figure~7-------------------------------------%
\begin{figure}
\begin{center}
\includegraphics[width=8.7cm]{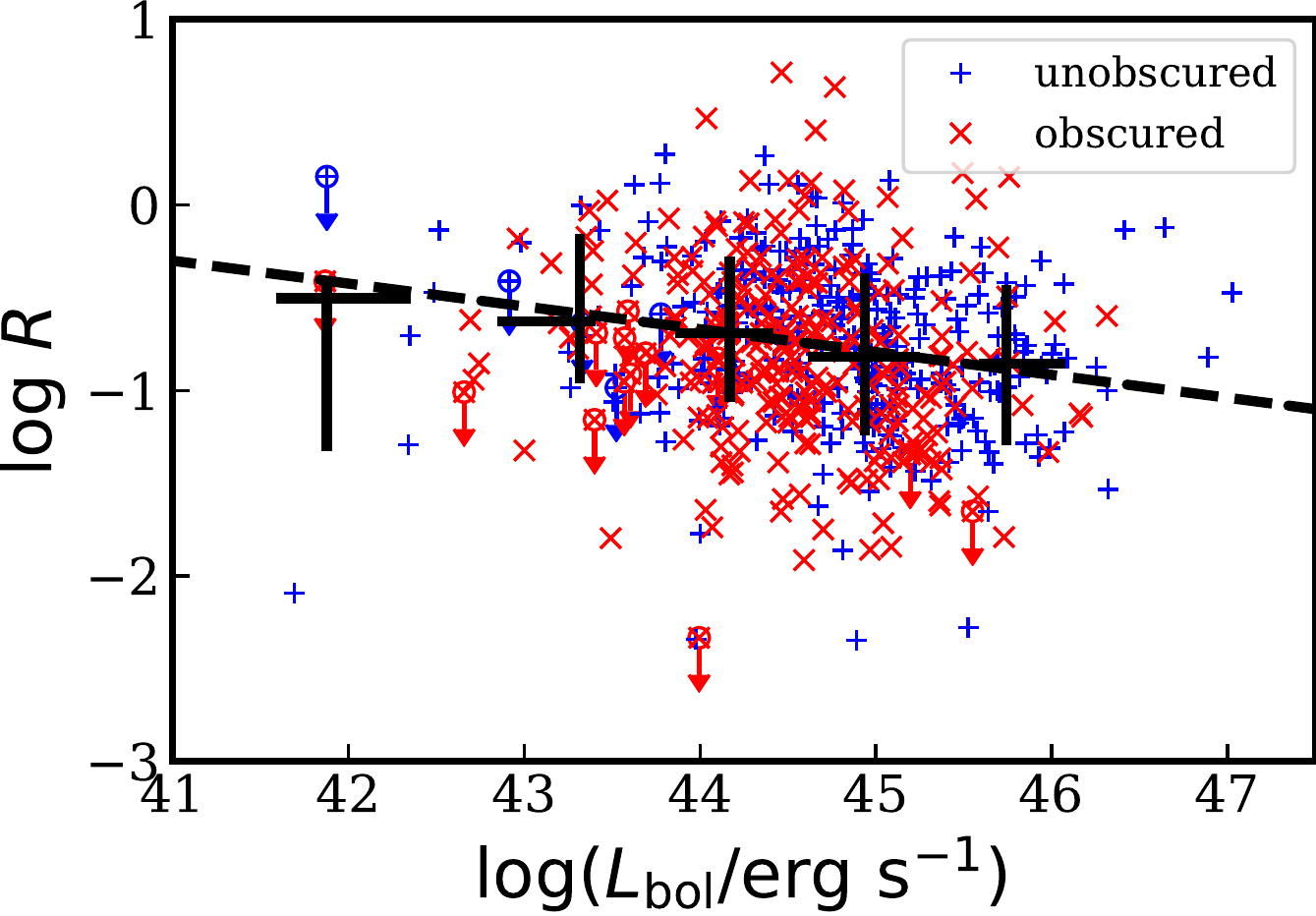}~
\caption{ $R=\ltorus/\lbol$  as a function of the bolometric luminosity.
The black crosses represent the median value of $R$ in each bin of the bolometric luminosity,
with the error bars showing the inter-percentage range with 68.2\% of the sample.
}\label{fig:RvsLbol}
\end{center}
\end{figure}
%------------------------------------------Figure~7-------------------------------------%

\subsection{Covering Factor of AGN as a Function of Bolometric Luminosity}\label{sec:CT}

The ratio of the AGN IR luminosity and the AGN bolometric luminosity 
($R=\lagnir/\lbol$)
has been interpreted as an indirect indicator of the dust covering factor $\cfdust$, since, for a given AGN luminosity,
 $\lagnir$ should be proportional to $\cfdust$ 
 \citep[$\lagnir \propto \cfdust \times \lbol$; ][]{mai07,tre08,eli12} .
Since the flux of the accretion disk cannot be directly measured for all the sources of our sample, 
we used $L_{14-150}$ to estimate the bolometric luminosity.
We apply a constant bolometric correction of $\lbol / L_{2-10}=20$,
which is equivalent to $\lbol / L_{14-150}= 8.47$ under the assumption of $\Gamma=1.8$,
which is the median value of the \textit{Swift}/BAT 70-month AGN sample \citep{ric16}.
We note that our main results do not change significantly when adopting different bolometric corrections, including luminosity-dependent ones
\citep{mar04}. We briefly discuss this in Appendix~\ref{sec:bolometric}.

%Added the definition of R based on the comment by Marko.
To calculate $R$, we follow in the same manner as \cite{sta16}.
We use the total IR AGN luminosity integrated over 1--1000~$\mu$m
($\ltorus$) instead of $\lagnir$ integrating the SED over 5--1000~$\mu$m.
This is because \cite{sta16} recommend to use the AGN SEDs including near-IR,
which sometimes contributes to the total IR luminosity with non-negligible level.
Since we do not have IR AGN template down to 1~$\mu$m,
we extrapolate the AGN template using the same spectral index
of $\alpha_1$ used at wavelength shorter than 19~$\mu$m.
Therefore, $R$ is calculated based on $R=\ltorus / \lbol $ in the following study.

 %R vs. Lbol
Figure~\ref{fig:RvsLbol} shows the relation between $R$ and the AGN bolometric luminosity.
The black dashed line represents the fit obtained using ASURV to account
for the sources with upper limit:
\begin{align}
\log R = (4.52 \pm 1.25) + (-0.12 \pm 0.03) \log \left( \frac{\lbol}{{\rm erg}~{\rm s}^{-1}} \right).
\end{align}

%------------------------------------------Figure~8-------------------------------------%
\begin{figure*}
\begin{center}
\includegraphics[width=0.75\linewidth]{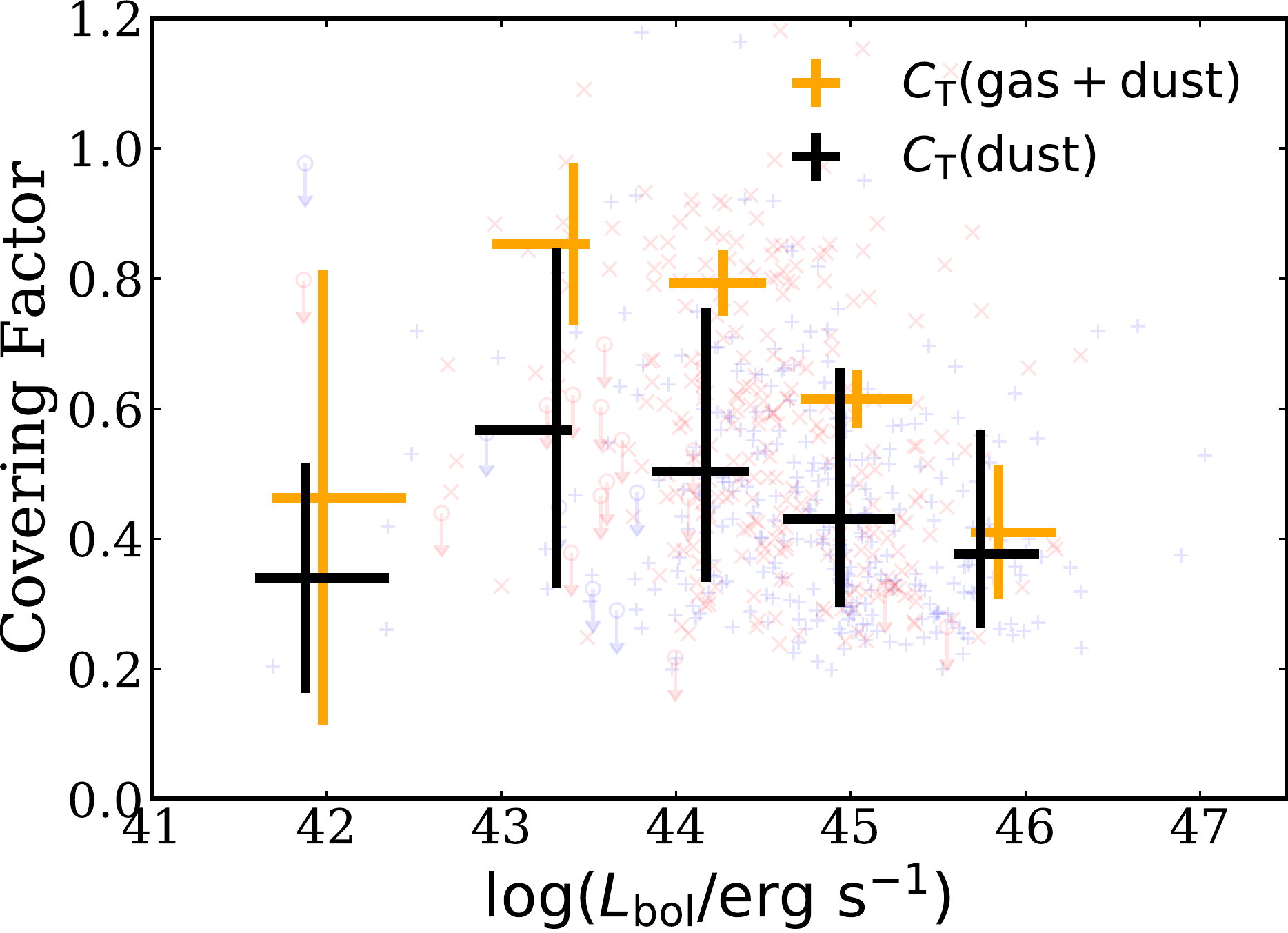}~
\caption{ 
The covering factor ($C_{\rm T}$) as a function of the bolometric luminosity.
 The dust covering factors $C_{\rm T}({\rm dust})$ are
 obtained from $R$ using the corrections
  reported in \cite{sta16}. The covering factor of gas and dust $C_{\rm T}({\rm gas+dust})$
  are obtained from the X-ray observations and the spectral fitting based on the
  obscured AGN fraction including the Compton-thick AGN \citep{ric15,ric16}.
The Compton-thick fraction is $f_{\rm CT}= 0.32$ for $\log (\lbol/{\rm erg}~{\rm s}^{-1}) < 43.5$ and $f_{\rm CT}= 0.21$ for $\log (\lbol/{\rm erg}~{\rm s}^{-1}) >43.5$.
 The orange crosses are shifted to the right by 0.1~dex for clarity.
}\label{fig:CFvsLbol}
\end{center}
\end{figure*}
%------------------------------------------Figure~8-------------------------------------%

This shows that $R$ is a very weak function of AGN bolometric luminosity.
However, $R$ does not always represent the actual $\cfdust$, because
the standard geometrically-thin and optically-thick disk emits radiation anisotropically \citep{net87,lus13}.
 Thus we also estimate $\cfdust$ exploiting the recent results of \cite{sta16}, 
 who computed the correction function between the covering factor ($\cfdust$) and $R$ using a clumpy two-phase medium
 with the sharp boundary between the dusty and dust-free environments.
They compute the $\cfdust$--$R$ relation
for a range of equatorial torus thickness ($\tau_{9.7}=3-10$).
We consider here the function for $\tau_{9.7}=3$:

{\small
\begin{equation}\label{eq:CFvsR}
\cfdust=
\begin{cases}
-0.178R^4+0.875R^3-1.487R^2\\ \hspace{25mm}+1.408R+0.192\ {\rm (type 1)}\\
2.039R^3-3.976R^2+2.765R+0.205\ {\rm (type 2)}.
\end{cases}
\end{equation}
}

We use the Equation~\ref{eq:CFvsR} 
for type-1/type-2 AGN to un/obscured AGN in this study.
According to \cite{sta16} the relations reported above are valid only for $R \le R_{\rm max}$, where $R_{\rm max}=1.3$ for unobscured AGN and $R_{\rm max} = 1.0$ for obscured AGN, so that we removed  five sources with $R \ge R_{\rm max}$
from the sample. Figure~\ref{fig:CFvsLbol} shows $\cfdust$ as a function of $L_{\rm bol}$.

%Description of CT(gas+dust)
Besides the dust covering factor $\cfdust$, we also calculate
the fraction of obscured AGN ($\log N_{\rm H}/\rm cm^{-2} \ge 22.0$), including the Compton-thick sources for each $\lbol$ bin as shown in Figure~\ref{fig:CFvsLbol}
(orange crosses). 
Since X-rays are absorbed by both gas and dust, 
the fraction of obscured AGN is a proxy of the covering factor of the obscuring material, and is sensitive to both gas and dust [$\cfgasdust$]
\footnote{
Although the dusty region also contains the gas, in this study
we use $\cfdust$ as the dusty covering factor which emits the IR emission
heated by AGN.  We then use $\cfgasdust$ as the covering area of gas
which is responsible for the X-ray absorption. This region includes
1) the dusty region defined by $\cfdust$ since the dusty region also includes
the gas, and 2) the dust-free region inside the sublimation radius,
but containing the neutral gas.}.
We follow the same approach to obtain $\cfgasdust$ of \cite{ric17a}.
The column density $N_{\rm H}$ for our sample is obtained through
the detailed X-ray spectral fitting using the follow-up X-ray observations \citep{ric16}.
In the X-ray fitting, both photoelectric absorption and Compton scattering are considered, and are listed in
Table~5 of \cite{ric16}. $\cfgasdust$ is defined as $\cfgasdust=f_{\rm Cthin}+
f_{\rm CT}$, where $f_{\rm Cthin}$ is the fraction of Compton-thin obscured AGN
 ($22 \le \log N_{\rm H}/\rm cm^{-2} < 24.0$) at each $\lbol$ bin,
 while the Compton-thick fraction is $f_{\rm CT}= 0.32$ for $\log (\lbol/{\rm erg}~{\rm s}^{-1}) < 43.5$ and $f_{\rm CT}= 0.21$ for $\log (\lbol/{\rm erg}~{\rm s}^{-1}) >43.5$ obtained from the intrinsic $N_{\rm H}$ distribution \citep{ric15}. The reason using the $f_{\rm CT}$ above is because even \textit{Swift}/BAT sources are unbiased for $N_{\rm H}< 10^{24}$~cm$^{-2}$, they can still be affected by obscuration for $N_{\rm H}> 10^{24}$~cm$^{-2}$.

%------------------------------------------Figure~CFvsLbol_sim-------------------------------------%
\begin{figure}
\begin{center}
\includegraphics[width=\linewidth]{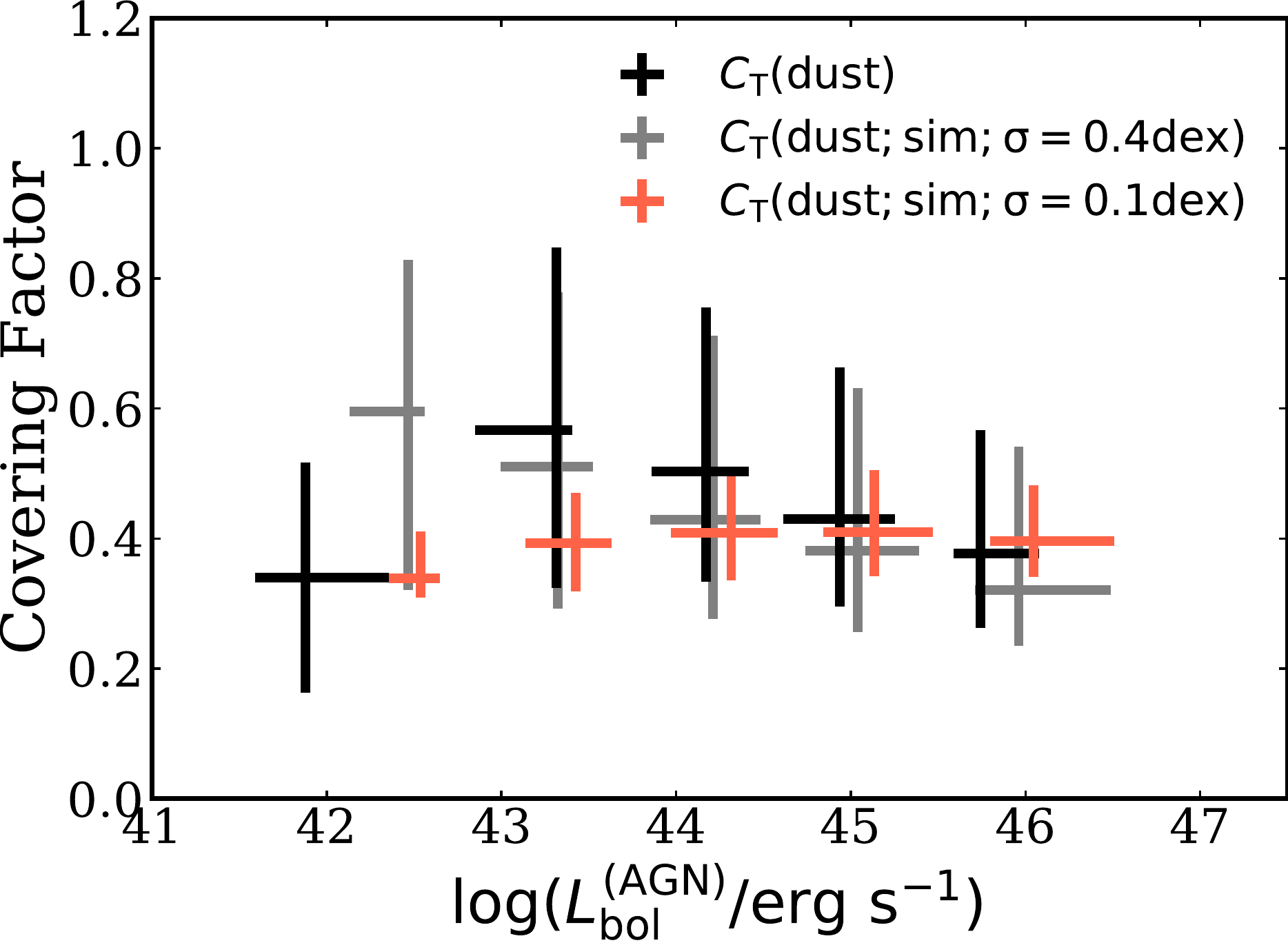}~
\caption{ 
The dust covering factor ($\cfdust$) as a function of the bolometric luminosity. 
 The simulated dust covering factor $\cfdustsim$ are
 obtained from the simulation using the random population following 
 the $\lagnir$--$L_{14-150}$ relation with scatter of $\sigma=0.4$ dex (gray
 crosses) and $\sigma=0.1$~dex (pink crosses).
 The gray/pink crosses are shifted to the right by 0.1/0.2~dex for clarity.
}\label{fig:CFvsLbol_sim}
\end{center}
\end{figure}
%------------------------------------------Figure~CFvsLbol_sim-------------------------------------%

\subsubsection{$\lbol$-dependent trend of $\cfdust$}\label{sec:Ldependent}
%CT(gas) shows the strong decreases, while CF(dust) shows a weak decrease.
Figure~\ref{fig:CFvsLbol} shows that both $\cfdust$ and $\cfgasdust$
seem to decrease as a function of AGN bolometric luminosity,
and, in the high luminosity end, $\cfgasdust$ finally converge into that of $\cfdust$.
%Luminosity dependence of CF(dust)
This luminosity-dependent trend of $C_{\rm T}$
has been observationally reported in multiple
wavelengths from IR \citep[e.g.,][]{mai07,alo11}, to the optical \citep{sim05},
and X-ray \citep{ued03,ued11,ued14,bec09,ric13}.

%However, there are some argument that L dependence is actually quite weak.
However, recent studies have found that
the luminosity-dependence of $\cfdust$
is actually really weak, 
and that the trend might even disppear
after considering some possible biases.
\cite{net16} argues that the reported $\cfdust$ would disappear by using different bolometric corrections.
\cite{sta16} also found that the luminosity-dependent trend
always mitigates after considering the anisotropy of the torus emission. 
A similar weak or non-significant
luminosity-dependent trend is reported in \cite{mat16},
and a more detailed review is given by \cite{net15}.

%Simulation of CF vs. Lbol using Daniel's script
In order to understand this trend in more detail,
we conduct simulations to assess the luminosity dependence of $\cfdust$.
We first generate two random populations of $L_{14-150}$ for
unobscured and obscured AGN for a sample of $10^4$ sources with the 
same number ratio as our parent sample
(unobscured/obscured=$300/287$; see Section~\ref{sec:sample}).
Each sample is generated based on our parent sample,
using Gaussian distribution with the median 
$\log (L_{14-150}/{\rm erg}~{\rm s}^{-1})$ of ($43.9, 43.6$) and the standard deviation of
of $(0.85~{\rm dex}, 0.67~{\rm dex})$ for unobscured and obscured AGN, respectively.
Then the distribution of $\ltorus$ is calculated under the assumption that
two populations follow the luminosity correlation of $\ltorus$--$L_{14-150}$
with a scatter of $\sigma=0.4$~dex, 
and finally the distribution of $\cfdust$ is computed in the same manner.
The result is shown in Figure~\ref{fig:CFvsLbol_sim}
that the computed $\cfdust$ distribution (gray cross bins)
roughly reproduces the luminosity-dependent trend of the black solid bins.
Next, we assume that all AGN should follow the luminosity correlation of 
$\ltorus$--$L_{14-150}$ and the intrinsic population should have
the narrower scatter; down to $\sigma=0.1$~dex.
The results are shown in pink in Figure~\ref{fig:CFvsLbol_sim},
showing that the luminosity-dependent trend disappears and
the binned $\cfdust$ is almost constant [$\cfdust \simeq 0.4$]
over the entire $\lbol$ range.
Therefore, we conclude that this seemingly luminosity-dependent trend 
can be produced purely by the scatter of the distribution, and our results
confirm the recent arguments that the luminosity-dependence 
of $\cfdust$ is actually really weak or absent.

\subsubsection{Relation between $\cfdust$ and $\cfgasdust$}\label{sec:cfdust_vs_cfgasdust}

%CF(gas+dust) always exceeds the CF(dust)
The other interesting result from the figure is that
the $\cfgasdust$ is always same or larger than the binned 
$\cfdust$ over the entire AGN luminosity range.
This relation still holds of $\cfdust \simeq 0.4 \leq \cfgasdust$
in our simulation as shown in Figure~\ref{fig:CFvsLbol_sim}.
This result suggests the presence of dust-free gas, 
possibly located in the broad line region (BLR),
 is responsible for part of the X-ray absorption \citep[see also][]{sch16}.
Observationally, using X-ray observations,
\cite{mar14} found evidence of occultation events in the X-rays, 
and the locations of those gas clumps are in the dust-free region
 or the inner edge of the dusty torus \citep[e.g.,][]{mai10,ris07,ris11}. 
In addition, \cite{min15} and \cite{gan15} have also suggested 
that the location of Fe K$\alpha$ line emitting material could be 
between the BLR and the dusty torus.
 Several studies have also proposed that the AGN gas disk inside the dust 
sublimation radius could significantly contribute to the observed column density in Compton-thick AGN,
since they are often found to have large inclination angles
\citep[e.g.,][]{dav15,mas16,ram17}.

 %Adding the sentence why SED Decomposition is important
 We also check whether the similar trend of
 $\cfdust \leq \cfgasdust$
 can be seen only using the MIR fluxes before the SED Decomposition. 
 This is discussed in Appendix~\ref{sec:CTobs}.

%------------------------------------------Figure~CFvsLbol_divtype-------------------------------------%
\begin{figure*}
\begin{center}
\includegraphics[width=0.5\linewidth]{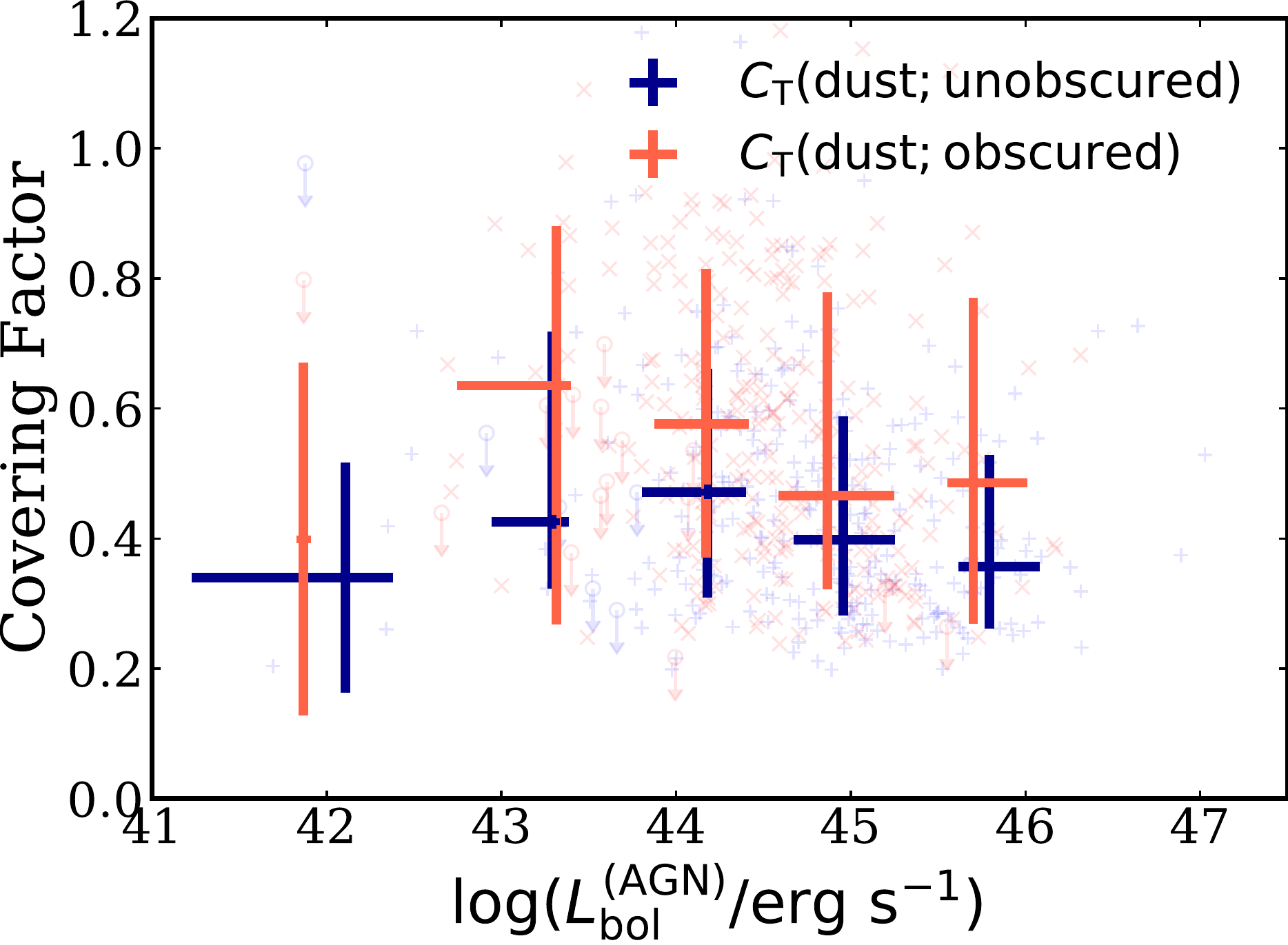}~
\includegraphics[width=0.5\linewidth]{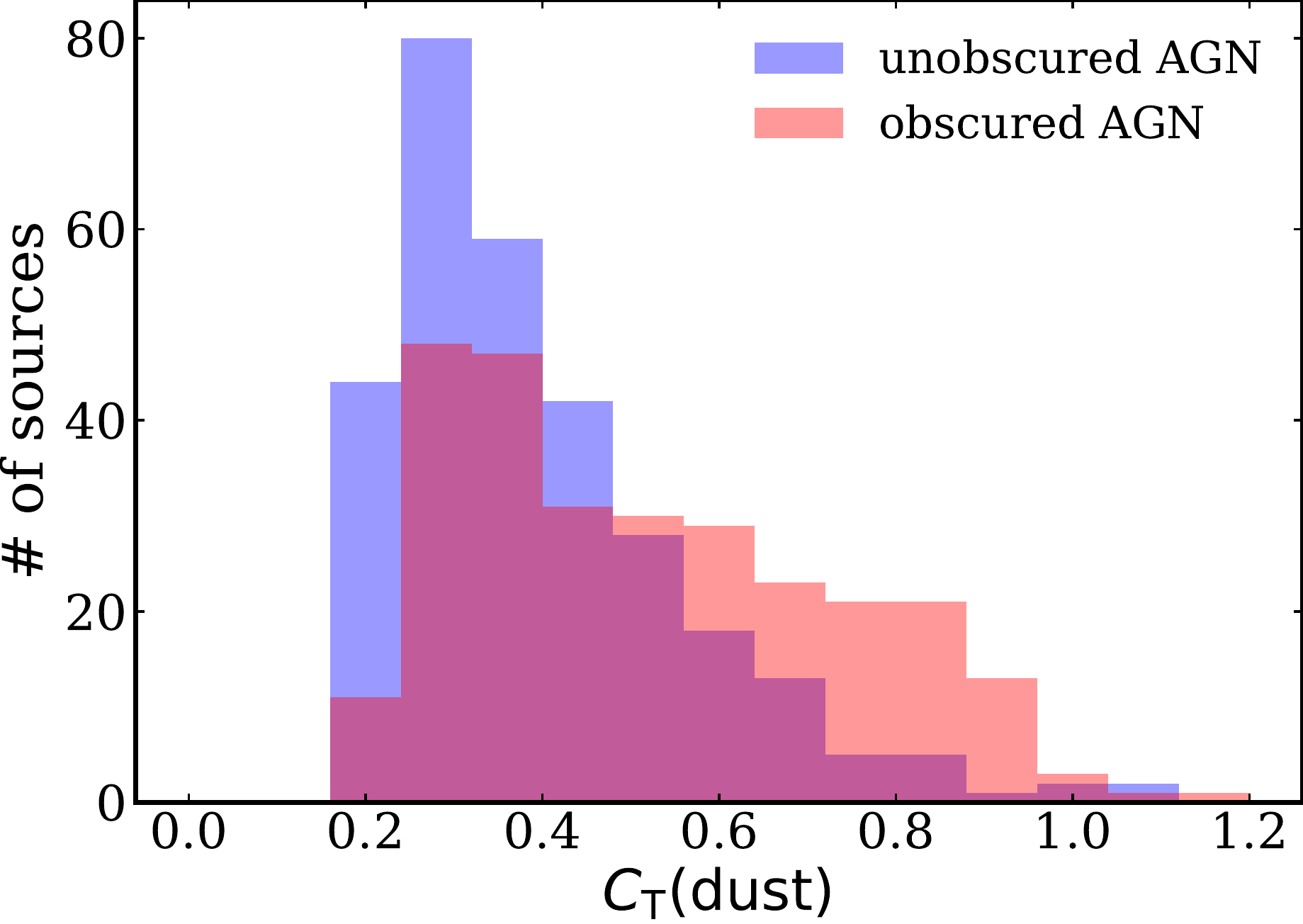}~
\caption{ 
(Left) The dust covering factor ($\cfdust$) of unobscured (blue) and obscured (red) AGN
as a function of the bolometric luminosity.
(Right) The distribution of $\cfdust$ of unobscured and obscured AGN.
}\label{fig:CFvsLbol_divtype}
\end{center}
\end{figure*}
%------------------------------------------Figure~CFvsLbol_divtype-------------------------------------%

%Both of CF(gas+dust) and CF(dust) shows the decline at the low-luminosity end?
Figure~\ref{fig:CFvsLbol} also shows that
both $\cfdust$ and $\cfgasdust$ seem to have a suggestive
peak at $\log L_{\rm bol} \simeq43$,
and they both seem decrease at lower luminosities. 
However, since the number of sample is limited in this bin range,
we cannot confirm the statistical significance of this trend
(see also the discussion in Appendix~\ref{sec:bolometric}).

\subsubsection{Comparison of $\cfdust$ between unobscured and obscured AGN}\label{sec:CF_AGNtype}
We compare here $\cfdust$ for the different subgroups of AGN.
The left panel of Figure~\ref{fig:CFvsLbol_divtype}
shows that the $\cfdust$ of unobscured (blue) and obscured (red) AGN
as a function of $\lbol$. Although the scatter is large, the binned 
$\cfdust$ of obscured AGN is always systematically higher than those of 
unobscured AGN.

The right panel of Figure~\ref{fig:CFvsLbol_divtype} shows
the distribution of $\cfdust$ for unobscured (blue) and obscured (red) AGN.
The $\cfdust$ distribution for unobscured AGN
is clustered at smaller values
of $\left< \cfdust \right>=0.41$, while the obscured AGN have a wider distribution of
$\cfdust$ reaching $\cfdust \simeq 1.0$.
We apply the KS test for these
two sample, and the null hypothesis 
$p$--value is $5.7 \times 10^{-8}$,
and the KS-statistics is 0.24,
suggesting that two distribution are significantly different.

One possible origin of the difference is that the smaller $\cfdust$ for unobscured AGN
could be due to larger $\lbol$. However, as discussed in Section~\ref{sec:Ldependent},
the luminosity dependence of $\cfdust$ is unlikely, and the KS test
shows that the distribution of $\cfdust$ for unobscured and obscured AGN
is statistically significant even in each $\lbol$ bin between 
$42.5< \log (\lbol/{\rm erg}~{\rm s}^{-1}) < 47$
with the $p$--value is $p<10^{-5}$ between
$42.5< \log (\lbol/{\rm erg}~{\rm s}^{-1}) < 45.5$
and $p=0.02$ between $45.5< \log (\lbol/{\rm erg}~{\rm s}^{-1}) < 47$.

Another possible interpretation of the difference is a consequence
of the selection of unobscured and obscured AGN.
Several authors argue that AGN classification depends on
the distribution of the $\cfdust$; unobscured AGN
would be preferentially observed from lower-$\cfdust$ AGN,
while obscured AGN from higher-$\cfdust$ AGN
\citep[e.g.,][]{ram11,eli12,ich15}.

%%%%%%%%%%%%CONCLUSIONS%%%%%%%%%%%%%%%%%%
\section{CONCLUSIONS}
We have constructed the IR (3--500~$\mu$m) SED for 587 nearby AGN 
detected in the 70~month \textit{Swift}/BAT all-sky survey.
Using this almost complete (587 out of 606; $94$\%) sample,
we have decomposed the IR (3--500~$\mu$m) SEDs into SB and AGN components.
The decomposition enabled us to estimate the AGN contribution
to the 12~$\mu$m ($\lagntwelve$), MIR ($\lagnmir$), and the total IR luminosity ($\lagnir$),
as well as the AGN luminosity contribution to the 12~$\mu$m ($\fagntwelve$), MIR ($\fagnmir$), and the total IR emission
 ($\fagnir$).
 Our results are summarized as follows:

\begin{enumerate}
\item The luminosity contribution of the AGN to the 12~$\mu$m, MIR, and total IR band flux increases with the 14--150~keV luminosity. 
For the most luminous sources, the AGN contribution is $\simeq 80$\% at the 12~$\mu$m, MIR, and $\simeq 50$\% at the total IR.
 
 \item We find nine pure IR-AGN whose IR emission is dominated
 by the AGN torus at least up to 90~\micron. 
 These pure IR-AGN could be a good candidates to create templates of the IR AGN SED, up to 90~\micron.
 Those sources could be easily selected using the color selection of 
 $f_{70~\mu{\rm m}}/f_{160~\mu{\rm m}}>1.0$ 
and $f_{22~\mu{\rm m}}/f_{70~\mu{\rm m}}>1.0$.

\item We find a good luminosity correlation between the MIR
 and ultra hard X-ray band over 5 orders of magnitude
 [$41< \log (L_{14-150}/ {\rm erg}~{\rm s}^{-1})<46$]. 
 Our slope is almost consistent with that obtained by studies carried out using high spatial resolution observations of nearby Seyfert galaxies, supporting
  our SED decomposition method, which would nicely estimate the 
 intrinsic MIR emission without the contamination of star-formation from the host galaxies.

 \item We find that the average of the covering factor of gas and dust inferred from X-ray observations always exceeds the one of dust torus covering factor, suggesting that the dust-free gas contributes to the absorption in the X-ray. 
 This gas could be associated inside the dust sublimation radius, 
 in agreement with previous observations based on X-ray occultation and spectral
 fitting studies of nearby AGN.

\item The luminosity-dependent trend of $\cfdust$ might
 be originated from the large scatter of the luminosity correlations between $\ltorus$
 and $L_{14-150}$, and the trend would be disappeared once the scatter is removed.

\item The obscured AGN tend to have larger $\cfdust$
than  unobscured AGN. This difference originates from
the AGN classification, which depends on the distribution of the obscuring material.

\end{enumerate}


%% Included in this acknowledgments section are examples of the
%% AASTeX hypertext markup commands. Use \url without the optional [HREF]
%% argument when you want to print the url directly in the text. Otherwise,
%% use either \url or \anchor, with the HREF as the first argument and the
%% text to be printed in the second.

\acknowledgments

We thank James Mullaney and Agnese Del Moro for providing the SB SED templates in this study,
 and Satoshi Takeshige for the technical discussion of IDL routine.
 We also thank Masatoshi Imanishi, Ryo Tazaki, and Daniel Asmus for fruitful discussions.
KI thanks the Department of Astronomy at Kyoto university, where a part of the research was conducted.
This study benefited from financial support from the Grant-in-Aid for JSPS fellow for young researchers (PD; KI), JSPS KAKENHI (18K13584; KI),
and JST grant ``Building of Consortia for the
Development of Human Resources in Science and Technology''
(KI).
CR acknowledges the CONICYT+PAI Convocatoria Nacional subvencion a instalacion en la academia convocatoria a\~{n}o 2017 PAI77170080.
FEB acknowledges support from CONICYT-Chile 
(Basal-CATA PFB-06/2007, FONDECYT Regular 1141218),
the Ministry of Economy, Development, and Tourism's Millennium Science
Initiative through grant IC120009, awarded to The Millennium Institute
of Astrophysics, MAS.
K.O. is an International Research Fellow of the Japan Society for the Promotion of Science (JSPS) (ID: P17321).

\clearpage
%------------------------------------------The Bibliography-------------------------------------%
%

%------------------------------------------TABLE~1-test-------------------------------------%
\onecolumngrid
\newpage

{\scriptsize
\begin{table}
\centering
\caption{Column descriptions for the IR catalog of \swift/BAT 70 Month AGN survey}
\begin{tabular}{lllll} \hline\hline \noalign{\smallskip}
Col. \# & Header Name & Format & Unit & Description \\
\noalign{\smallskip} \hline \noalign{\smallskip}
1 & objID & string & -- & \swift/BAT ID as shown in \cite{bau13} \\
2 & ctpt1 & string & -- & optical counterpart name \\
3 & $z$ & float & --  & redshift \\
4 & NH\_log & float & -- & logarithmic column density ($\log N_{\rm H}/{\rm cm}^{-2}$) \\
5 & lbat\_log & float & -- & absorption corrected logarithmic 14--150~keV luminosity ($\log L_{14-150}/{\rm erg}~{\rm s}^{-1}$) \\
6 & lbol\_const\_log & float & -- & logarithmic bolometric AGN luminosity 
($\log (\lbol/{\rm erg}~{\rm s}^{-1})$) \\
7 & lbol\_log & float & -- & logarithmic bolometric AGN luminosity 
($\log (L_{\rm bol}^{(\rm M04)}/{\rm erg}~{\rm s}^{-1})$) using \cite{mar04} \\
8 (9) & fnu3p4\_(err)\_fqualmod & float & Jy & 3.4~\micron~profile-fitting flux density (error) obtained from \wise \\
10 (11) & fnu4p6\_(err)\_fqualmod & float & Jy & 4.6~\micron~profile-fitting flux density (error) obtained from \wise \\
12 (13) & fnu9a\_(err)\_fqualmod & float & Jy & 9.0~\micron~flux density (error) obtained from \akari/IRC \\
14 (15) & fnu12wipf\_(err)\_fqualmod & float & Jy & 12~\micron~flux density (error)  \\
16 & fnu12wipcatalog & string & -- & reference catalogs for 12~\micron: W=\wise, Ip=\iras/PSC, If=\iras/FSC \\
17 (18) & fnu18a\_(err)\_fqualmod & float & Jy & 18.0~\micron~flux density (error) obtained from \akari \\
19 (20) & fnu22w\_(err)\_fqualmod & float & Jy & 22~\micron~profile-fitting flux density (error) obtained from \wise \\
21 (22) & fnu25ipf\_(err)\_fqualmod & float & Jy & 25~\micron~flux density (error) \\
23 & fnu25ipfcatalog & string & -- & reference catalogs for 25~\micron: Ip=\iras/PSC, If=\iras/FSC \\
24 (25) & fnu60ipf\_(err)\_fqualmod & float & Jy & 60~\micron~flux density (error)  \\
26 & fnu60ipfcatalog & string & -- & reference catalogs for 60~\micron: Ip=\iras/PSC, If=\iras/FSC \\
27 (28) & fnu65a\_(err)\_fqualmod & float & Jy & 65~\micron~flux density (error) obtained from \akari/FIS \\
29 (30) & fnu70p\_(err)\_fqualmod & float & Jy & 70~\micron~flux density (error) obtained from \herschel/PACS \\
31 (32) & fnu90a\_(err)\_fqualmod & float & Jy & 90~\micron~flux density (error) obtained from \akari/FIS \\
33 (34) & fnu100ipf\_(err)\_fqualmod & float & Jy & 100~\micron~flux density (error) \\
35 & fnu100ipfcatalog & string & -- & reference catalogs for 100~\micron: Ip=\iras/PSC, If=\iras/FSC \\
36 (37) & fnu140a\_(err)\_fqualmod & float & Jy & 140~\micron~flux density (error) obtained from \akari/FIS \\
38 (39) & fnu160pa\_(err)\_fqualmod & float & Jy & 160~\micron~flux density (error)  \\
40 & fnu160pacatalog & string & -- & reference catalogs for 160~\micron: P=\herschel/PACS, A=\akari/FIS \\
41 (42) & fnu250s\_(err)\_fqualmod & float & Jy & 250~\micron~flux density (error) obtained from \herschel/SPIRE \\
43 (44) & fnu350s\_(err)\_fqualmod & float & Jy & 350~\micron~flux density (error) obtained from \herschel/SPIRE \\
45 (46) & fnu500s\_(err)\_fqualmod & float & Jy & 500~\micron~flux density (error) obtained from \herschel/SPIRE \\ \hline
47 & l12\_AGN\_afSta15\_log & float & -- & Logarithmic decomposed 12~$\mu$m AGN luminosity $\log (\lagntwelve/{\rm erg}~{\rm s}^{-1})$ \\
48 & lMIR\_AGN\_afSta15\_log & float & -- & Logarithmic decomposed MIR AGN luminosity $\log (\lagnmir/{\rm erg}~{\rm s}^{-1})$ \\
49 & lIR\_AGN\_afSta15\_log & float & -- & Logarithmic decomposed total IR luminosity $\log (\lagnir/{\rm erg}~{\rm s}^{-1})$ \\
50 & l12AGNratio\_afSta15 & float & -- & $\fagntwelve$ \\
51 & AGNpercentage\_MIR\_afSta15 & float & -- & $\fagnmir$ \\
52 & AGNpercentage\_afSta15 & float & -- & $\fagnir$ \\
53 & flag\_upperlimit & int & -- & Flag of AGN: detection ($=0$), upper-limit ($=1$), and lower-limit ($=-1$)\\
54 & R\_Sta16\_afSta15\_log & float & -- & $\log R= \log (\lagnir/\lbol)$\\
55 & CF\_Sta16\_tau9p7eq3\_afSta15 & float & -- & $\cfdust$\\ 
56 & SBtemplate\_afSta15 & string & -- & SB template used for the SED fitting this study: SB1--SB5\\
\noalign{\smallskip} \hline \noalign{\smallskip}
\end{tabular}
\begin{minipage}[l]{0.76\textwidth}
\footnotesize
\textbf{Notes.}
The detail of the selection of the flux is compiled in Section~\ref{sec:sample}. 
The full catalog is available as a machine readable electronic table.
\end{minipage}
\label{tab:IRcatalog}
\end{table}
}
%------------------------------------------TABLE~1-------------------------------------%

%------------------------------------------tab:Eq-------------------------------------%
\setlength{\topmargin}{3.0cm}
\floattable
\renewcommand{\arraystretch}{1.1}
\begin{deluxetable}{cccccccc}
\thispagestyle{empty}
  \tablecaption{Equations of the correlation in this study}\label{tab:Equation}
  \tablehead{
      \colhead{(1)} & \colhead{(2)} & \colhead{(3)} & \colhead{(4)} & \colhead{(5)}\\
       \colhead{Y} & \colhead{X} & \colhead{$a$} & \colhead{$b$} & \colhead{Reference}
     }
    \startdata
\thispagestyle{empty}
%12um vs. Lbat
$\log \left( \frac{\lagntwelve}{10^{43}~{\rm erg/s}} \right)$ & 
$\log \left( \frac{L_{\rm 14-150}}{10^{43}~{\rm erg/s}} \right)$ &
$-0.24 \pm 0.03$ & $1.08 \pm 0.03$ & Section~\ref{sect:LxvsLIR}\\
%LMIR vs. Lbat
$\log \left( \frac{\lagnmir}{10^{43}~{\rm erg/s}} \right)$ & 
$\log \left( \frac{L_{\rm 14-150}}{10^{43}~{\rm erg/s}} \right)$ &
$-0.05 \pm 0.03$ & $1.06 \pm 0.03$ & Section~\ref{sect:LxvsLIR}\\
%R vs. Lbol
$\log R$ & $\log \left( \frac{\lbol}{{\rm erg}~{\rm s}^{-1}} \right)$ &
$4.52 \pm 1.25$ & $-0.12 \pm 0.03$ & Section~\ref{sec:CT}
    \enddata
    \thispagestyle{empty}
\tablenotetext{}{
Notes.--- Correlation properties between two physical values.
Columns: (1) Y variable; (2) X variable; (3) regression intercept ($a$)
and its 1$\sigma$ uncertainty; (4) slope ($b$) and its 1$\sigma$ uncertainty.
Equation is represented as $Y=a+bX$; (5) Reference of the details on each Equation.
}\label{tab:Equation}
\end{deluxetable}
\setlength{\topmargin}{0in}
%------------------------------------------tab:Eq-------------------------------------%

%\clearpage

 %------------------------------------------Figure~A14vsDecomp-------------------------------------%
\begin{figure*}
\begin{center}
\includegraphics[width=\linewidth]{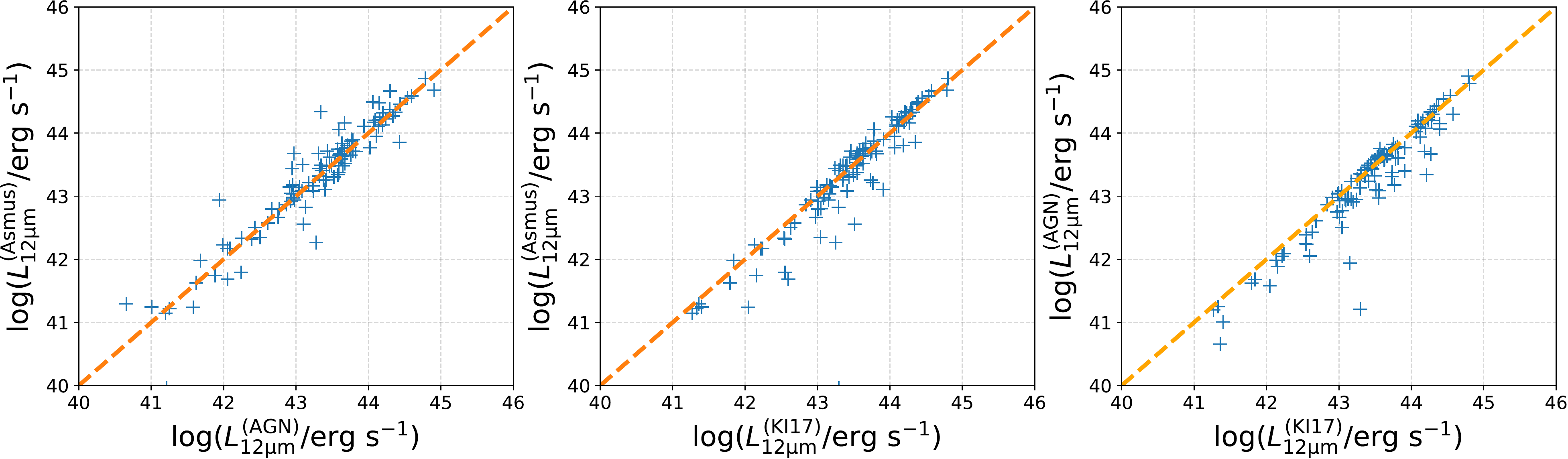}\\
\includegraphics[width=\linewidth]{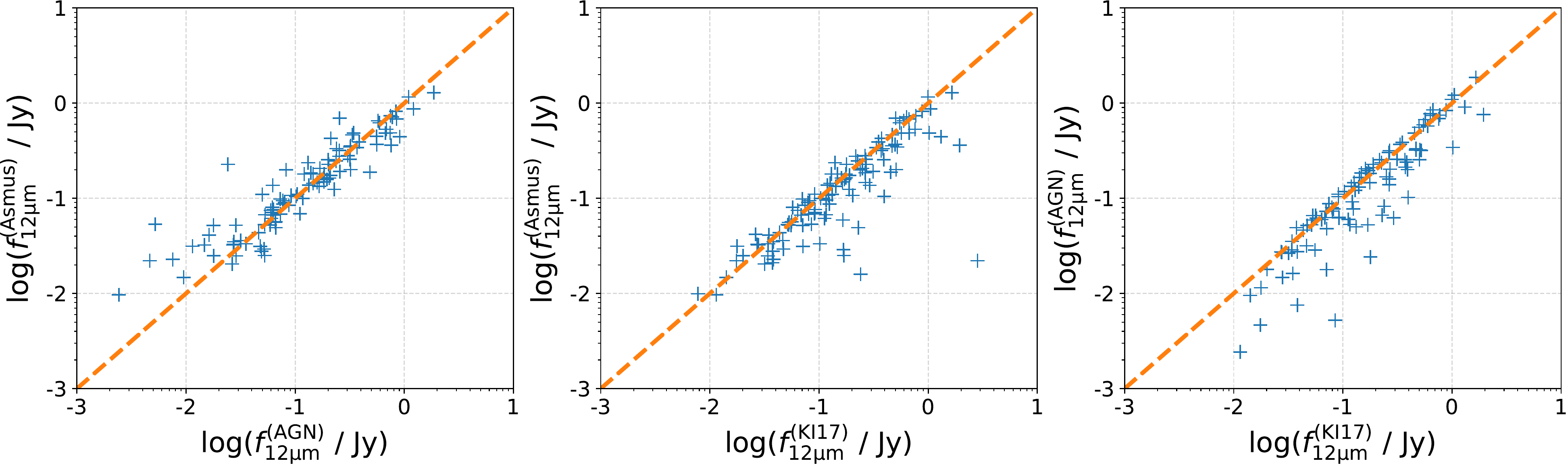}~
\caption{
(Top) Scatter plot of the 12~$\mu$m luminosities obtained from
high spatial resolution MIR observations \citep[$L_{\rm 12{\mu}m}^{(\rm Asmus)}$;][]{asm14,asm15}, this study after the SED decomposition ($\lagntwelve$),
 and before the SED decomposition \citep[$L_{\rm 12{\mu}m}^{(\rm KI17)}$;][]{ich17}.
 The blue cross represents the individual sources, and the orange dashed line represents
 the 1:1 relation.
 Each panel shows the luminosity relation between 
 $L_{\rm 12{\mu}m}^{(\rm Asmus)}$ and $\lagntwelve$ (Left),
 $L_{\rm 12{\mu}m}^{(\rm Asmus)}$ and $L_{\rm 12{\mu}m}^{(\rm KI17)}$ (Middle),
 $\lagntwelve$ and $L_{\rm 12{\mu}m}^{(\rm KI17)}$ (Right). 
 (Bottom) Same plots as top but for 12~$\mu$m flux densities.
}\label{fig:A14vsDecomp}
\end{center}
\end{figure*}
%------------------------------------------Figure~LxvsLIR-------------------------------------%

\appendix

\section{Comparison with the studies from the literature}\label{sec:CompLiterature}
\subsection{Comparison with the high spatial resolution flux obtained with ground-based 8m class telescopes}\label{sec:CompAsm14}
Here we compare the results in this study with the high spatial resolution observations
by \cite{asm14,asm15}.
Out of 122 high spatial resolution sources, we found 112 sources also used in this
study. The remaining 10 sources were not found because they are located in the
low galactic latitude of $|b|<10^{\circ}$, where we initially removed from the parent 
sample as discussed in \cite{ich17}.

The top left panel of Figure~\ref{fig:A14vsDecomp} shows
the 12~$\mu$m luminosity correlation between 
the high spatial resolution MIR observations \citep[$L_{\rm 12{\mu}m}^{(\rm Asmus)}$;][]{asm14,asm15} and this study after the SED decomposition ($\lagntwelve$). 
The figure clearly shows that our decomposition method successfully follows
the one-by-one relation with the high spatial resolution observations down to $\log (\lagntwelve/{\rm erg}\ {\rm s}^{-1}) \simeq 41.0$.
The average of two parameters of $\left<\log L_{\rm 12{\mu}m}^{(\rm Asmus)}/
\lagntwelve \right> = 0.05$. The standard deviation is $\sigma=0.36$.

The top middle and right panels of Figure~\ref{fig:A14vsDecomp}
shows the luminosity relation between
$L_{\rm 12{\mu}m}^{(\rm Asmus)}$, $\lagntwelve$,
and the low-resolution 12~$\mu$m luminosity before the SED decomposition
$(L_{\rm 12{\mu}m}^{(\rm KI17)})$, which is taken from \cite{ich17}.
Both panels show that the points are distributed equal or below the
one-by-one relations and suggestive of the contamination of the
host galaxy component in 
$L_{\rm 12{\mu}m}^{(\rm KI17)}$.
The mean and standard devition is $\left<\log L_{\rm 12{\mu}m}^{(\rm Asmus)}/
L_{\rm 12{\mu}m}^{(\rm KI17)} \right> = -0.10 \pm 0.43$. 
This shows that the correlation between
$L_{\rm 12{\mu}m}^{(\rm Asmus)}$ and $\lagntwelve$ is tighter 
than that of $L_{\rm 12{\mu}m}^{(\rm Asmus)}$ and $L_{\rm 12{\mu}m}^{(\rm KI17)}$, indicating that our decomposition method nicely reduces the
contamination in the 12~$\mu$m band from the host galaxies.

The bottom panels of Figure~\ref{fig:A14vsDecomp} show the
same relation as those in the top panels, but for
12~$\mu$m flux densities. All three panels also show the similar trend as
the luminosity relations. One notable difference is that the flux density
of the high spatial resolution observation ($f_{\rm 12 {\mu}m}^{\rm (Asmus)}$)
shows a decline of the number of sources at around 
$f_{\rm 12 {\mu}m}^{\rm (Asmus)} \simeq 10^{-2}$~Jy.
This is almost consistent with the lower-bound of the flux density
observable with ground-based 8m class telescopes with significant signal-to-noise ratio \citep{asm14}. 
Our study can explore flux densities down to
$10^{-3}$~Jy, which is equivalent to the detection limit of the \textit{WISE} W3 ($12$~$\mu$m) band.
This is one of the advantages of the SED decomposition method using low-resolution,
but sensitive space IR satellites compared to the ground based studies.

\subsection{Comparison with different models from the literature}
In this appendix we briefly compare the IR AGN luminosity obtained in this study
and the ones obtained in \cite{shi17}.
They applied a different IR SED model
to the IR dataset, which is similar with ours, obtained from
the \textit{Herschel} observations in
the \textit{Swift}/BAT 58-month AGN catalog
to study mainly the global star-forming properties in the host galaxies.
Instead of using the AGN/host galaxy templates, 
they provided a function of hot dust and the host galaxy respectively
by following \cite{cas12} and their functions are given by
\begin{equation}
f(\nu)= N_{\rm pl}\left( \frac{\nu}{\nu_{\rm c}} \right) e^{-(\nu_{\rm c}/\nu)^2}
+S_{\rm MBB}(\nu, M_{\rm dust}, T_{\rm dust}),
\end{equation}
where the first section stands for the AGN component with the normalization
$N_{\rm pl}$, cut-off frequency $\nu_{\rm c}$, and the second section
represents the host galaxy component of a single modified black body
with the parameter of dust mass $M_{\rm dust}$ and the dust temperature $T_{\rm dust}$. The fitting method used in their study is also different with ours.
They use a Bayesian framework with a Markov chain Monte Carlo to obtain
the posterior probability distribution function, and then use the median
to obtain the best fitted parameters.
Out of 307 sources in their sample, 204 sources have at least one
\textit{Herschel} detections and the reliable fitting quality (\texttt{lir\_agn\_flag}$=0$).
After the cross-matching with our sample, we found 180 sources in common.
Again, the removed 24 sources are located in the low galactic latitude of $|b|< 10^{\circ}$.

\begin{figure*}
\begin{center}
\includegraphics[width=0.7\linewidth]{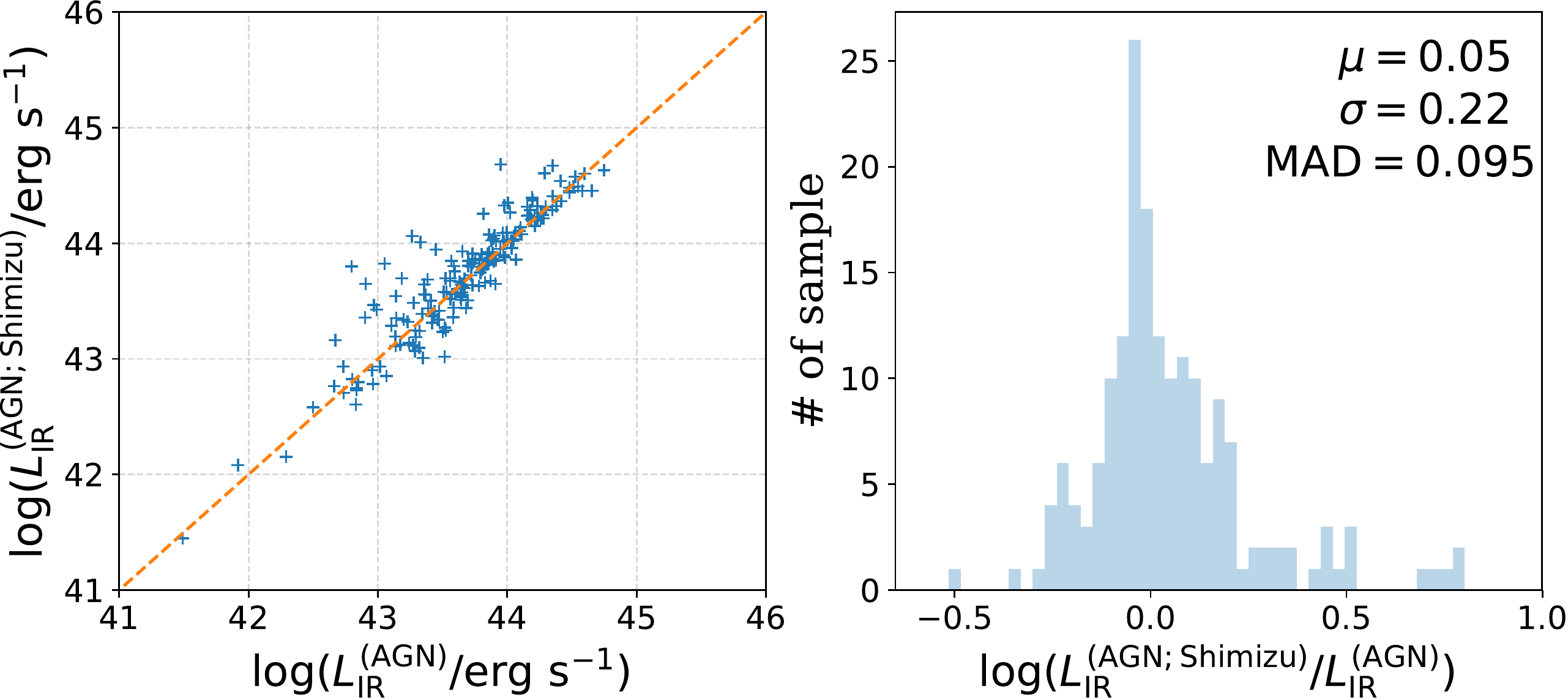}~
\caption{
(Left) Scatter plot of total IR AGN luminosities obtained from
\cite{shi17} ($L_{\rm IR}^{(\rm AGN; Shimizu)}$) 
and ones obtained from this study ($\lagnir$).
 The blue cross represents the individual sources, and the orange dashed line represents the 1:1 relation.
(Right) Histogram of $r=\log ( L_{\rm IR}^{(\rm AGN; Shimizu)}/ \lagnir )$.
The mean $\mu$, standard deviation $\sigma$,
and median absolute deviation (MAD) of $r$
 are also shown in the panel.
}\label{fig:S17vsDecomp}
\end{center}
\end{figure*}

Since \cite{shi17} do not provide any 12~$\mu$m AGN flux or luminosity,
 we compare the total IR AGN luminosity obtained from their AGN component.
The left panel of Figure~\ref{fig:S17vsDecomp} shows the
correlation between the IR AGN luminosities obtained from \cite{shi17}
($L_{\rm IR}^{(\rm AGN; Shimizu)}$) and the ones from this study.
We find a good luminosity correlation between $L_{\rm IR}^{(\rm AGN; Shimizu)}$ and $\lagnir$.
The Spearman's ran coefficient is 0.91,
and null hypothesis probability is $P=4.9 \times 10^{-69}$,
suggesting the correlation is significant.
The average of the distribution of $r=\log ( L_{\rm IR}^{(\rm AGN; Shimizu)}/ \lagnir )$ is also shown in the right panel of Figure~\ref{fig:S17vsDecomp}.
We do not find any systematic offset between the two methods 
$(\mu=0.05)$ with a standard deviation of $\sigma=0.22$~dex.
Since there are several outliers with $\log ( L_{\rm IR}^{(\rm AGN; Shimizu)}/ \lagnir ) > 0.3$, we also compute the median absolute deviation (MAD)
and the value is MAD$=0.095$~dex, which is smaller by a factor of two
than the standard deviation.
As already mentioned in \cite{shi17}, 
their model allows the power-law component to extend to longer wavelengths,
which would return slightly larger AGN luminosities with
$\log ( L_{\rm IR}^{(\rm AGN; Shimizu)}/ \lagnir ) > 0.3$ for some cases.
Those sources are actually seen in Figure~\ref{fig:S17vsDecomp}
but only for a few percentages of the sample. Thus,
we conclude that, although the different fitting methods and the template, each model returns the consensus results for the estimation of the IR AGN luminosities.

 %------------------------------------------Figure~S17vsDecomp-------------------------------------%
%------------------------------------------Figure~S17vsDecomp-------------------------------------%

%------------------------------------------Figure~CFvsLbol_M04-------------------------------------%
\begin{figure*}
\begin{center}
\includegraphics[width=0.5\linewidth]{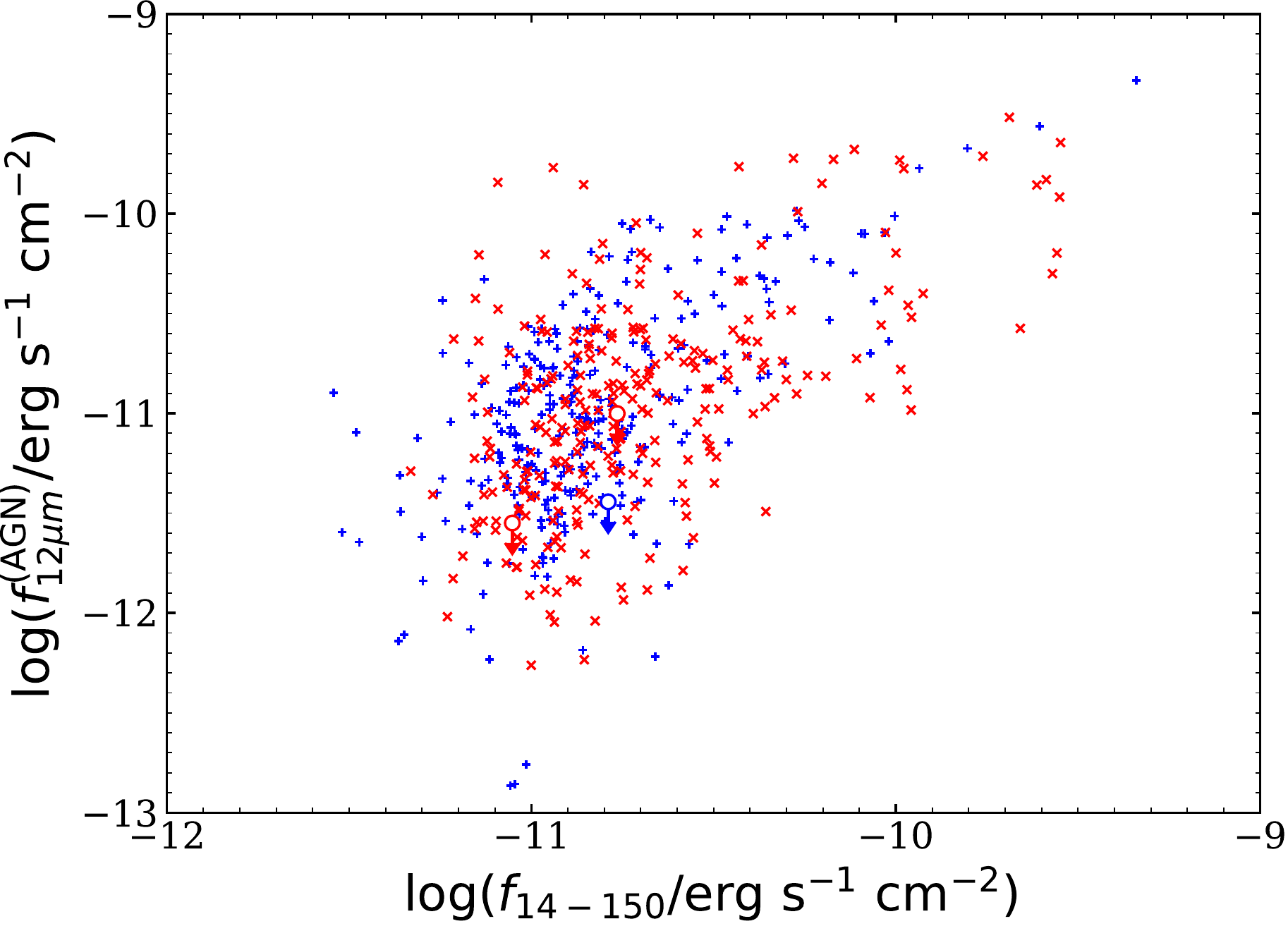}~
\includegraphics[width=0.5\linewidth]{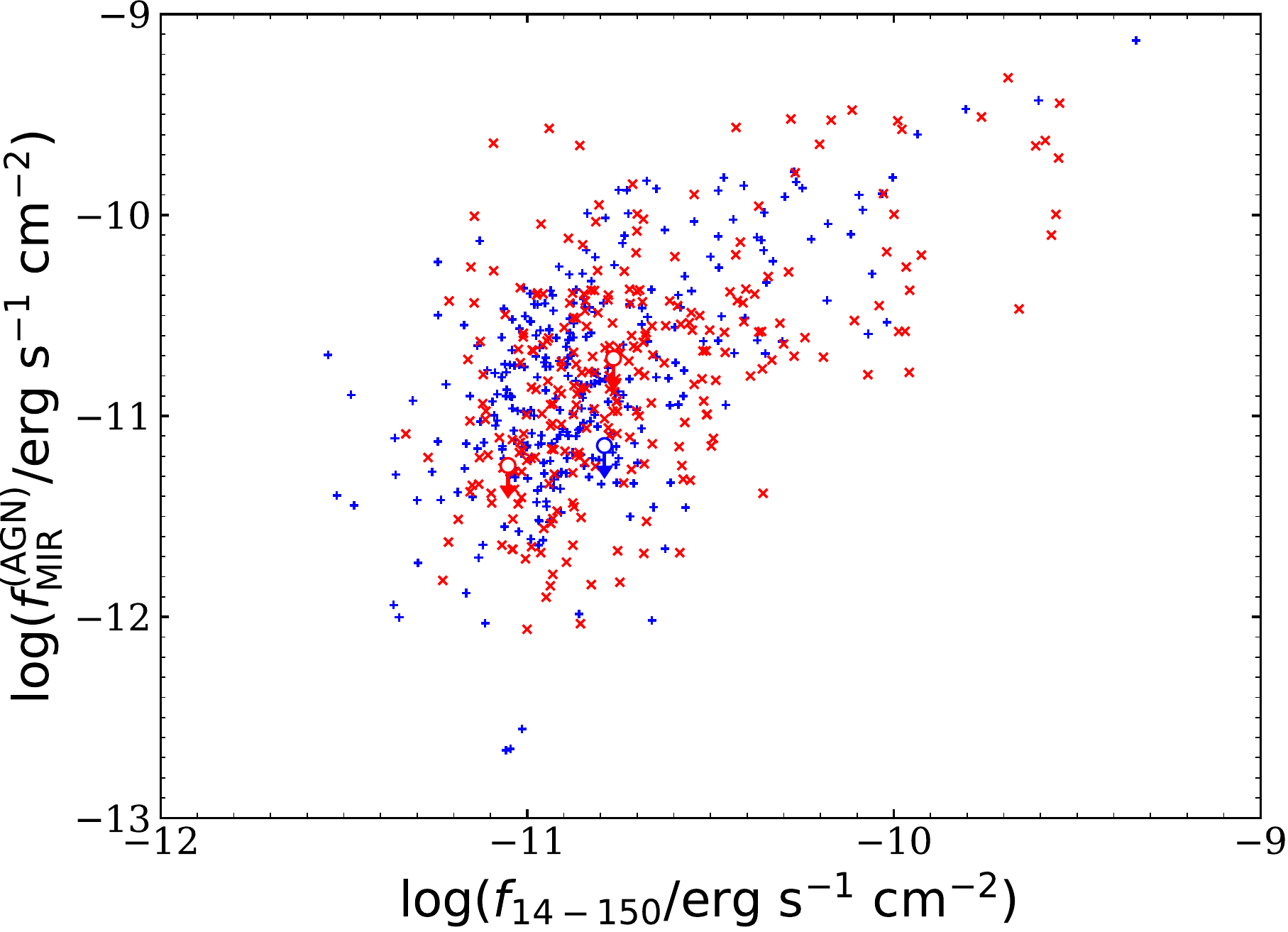}~
\caption{ 
Correlation between the fluxes at 12~$\mu$m, MIR band and 14--150 keV. 
Blue and red cross represents unobscured and
obscured AGN, respectively.
}\label{fig:fIRvsfbat}
\end{center}
\end{figure*}
%------------------------------------------Figure~CFvsLbol_M04-------------------------------------%

\section{Flux Correlation Between $12$~$\mu$m, 
MIR, and 14--150 keV bands}\label{sec:fIRvsfbat}

Figure~\ref{fig:fIRvsfbat} shows the flux correlation between
the AGN 12~$\mu$m, MIR, and 14--150~keV bands,
showing a clear correlation between the two bands even
in the flux-flux plane. 
The Spearman's ran coefficient is 0.43 and
the null hypothetical probability is $P=10^{-28}$ for 
both flux-flux correlations, suggesting the correlation is significant.
The slope of $b=1.48$ for the AGN 12~$\mu$m band
and $b=1.49$ for the AGN MIR band, respectively.
As we discussed in \cite{ich17}, there is a clear decline
of the number of sources at $f_{14-150}< 10^{-11}$~erg~s$^{-1}$~cm$^{-2}$,
while MIR flux can go down to $3 \times 10^{-13}$~erg~s$^{-1}$~cm$^{-2}$, which is
the typical MIR band detection limit.
This trend suggests that the sample is limited by the
X-ray flux detection limit.

%------------------------------------------Figure~CFvsLbol_M04-------------------------------------%
\begin{figure*}
\begin{center}
\includegraphics[width=0.5\linewidth]
{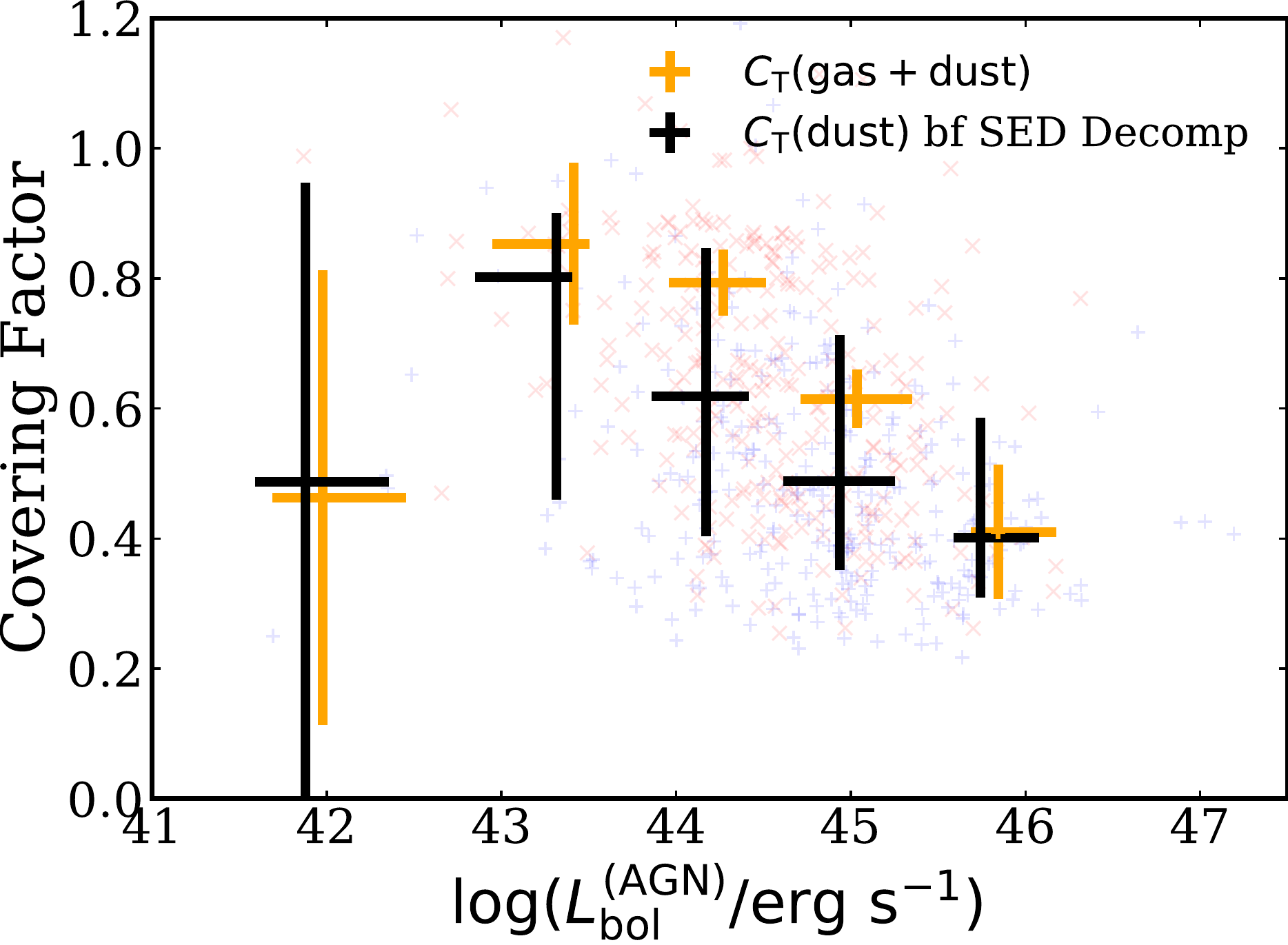}~
\includegraphics[width=0.5\linewidth]{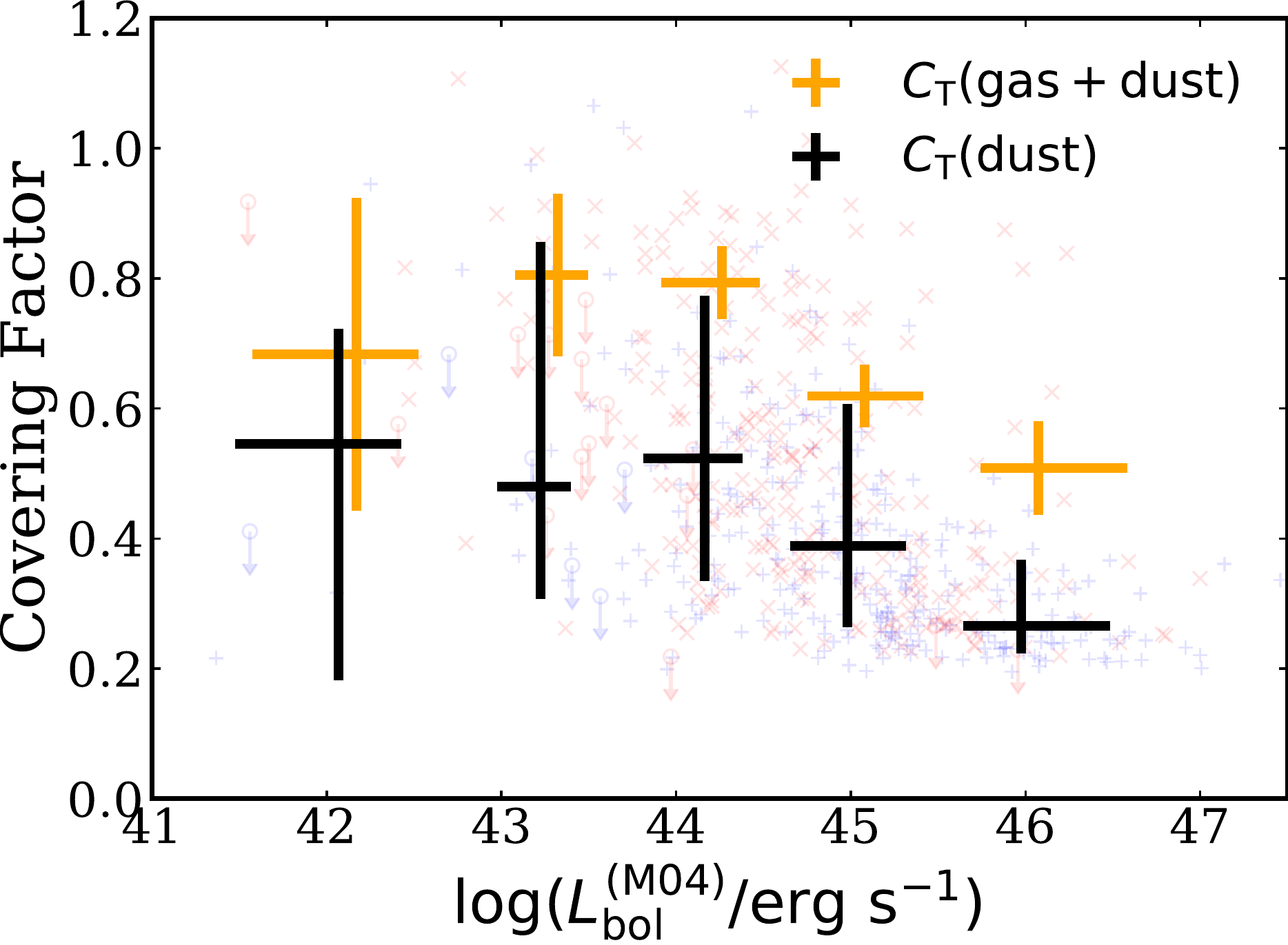}~
\caption{ 
Same as Figure~\ref{fig:CFvsLbol}, but using the different estimation for the values.
(Left) the covering factors used here are based on the estimation
using the observed 12~$\mu$m luminosities in \cite{ich17} before the IR SED decomposition.
(Right) the bolometric corrections used are dependent on the
bolometric luminosity \citep[$L_{\rm bol}^{(\rm M04)}$;][]{mar04},
not the constant bolometric correction.
}\label{fig:CFvsLbol_M04}
\end{center}
\end{figure*}
%------------------------------------------Figure~CFvsLbol_M04-------------------------------------%

\section{Comparison of $C_{\rm T}$ and $\lbol$ Relation Using Different Values}
\subsection{$\cfdust$ Estimated from the Observed 12~$\mu$m Luminosity}\label{sec:CTobs}
It is important to check whether the same result in Figure~\ref{fig:CFvsLbol}
is obtained using the MIR fluxes without host galaxy subtraction.
To achieve this, we estimate the total IR AGN luminosity
by assuming 
that the observed 12~$\mu$m luminosity originates from the AGN emission.
Then we use the conversion factor of $\ltorus / \lagntwelve=2.77$
estimated from the AGN template in this study. 
The calculation of the $R$, and then $\cfdust$
is performed in the same manner as we discussed in Section~\ref{sec:CT}.
The left panel of Figure~\ref{fig:CFvsLbol_M04} shows the relation
of $C_{\rm T}$ and $\lbol$ using the $\cfdust$ estimated above.
It clearly shows that while the result of $\cfdust < \cfgasdust$ holds between 
$43.5< \log \lbol < 45.5$, $\cfdust$ becomes almost equal to $\cfgasdust$ at the luminosity bin of $42.5 < \log \lbol < 43.5$, which is not seen in Figure~\ref{fig:CFvsLbol}.
We also apply the KS-test between $\cfdust$ and $\cfgasdust$ for each 
$\lbol$ luminosity bin. In order to apply the KS-test, we make a Gaussian 
distribution of $\cfgasdust$ in which the central value is the average of $\cfgasdust$ and the 1$\sigma$ is the standard deviation of $\cfgasdust$, and the number of sources 
are same as $\cfdust$ in the same $\lbol$ bin.
As a result, we find a significant difference for the luminosity bins 
between $43.5< \log \lbol < 45.5$ with $p$-values of $p<10^{-30}$, while
the clear significance is not obtained at the luminosity bin of $ \log \lbol < 43.5$
($p>0.5$) and $45.5 < \log \lbol $ ($p=0.26$).
This difference would originate from the flux subtraction after the
SED decomposition especially at the lower AGN luminosity end,
suggesting their importance and effect to estimate the dust covering factor.

\subsection{Dependence of the Bolometric Corrections}\label{sec:bolometric}
Here we summarize whether different bolometric correction can affect
the relation shown in Figure~\ref{fig:CFvsLbol}.
In this study, following the method used in \cite{ric17a},
we use the constant bolometric correction of $\lbol / L_{14-150}=8.47$,
which is based on $\lbol / L_{2-10}=20$ under
the assumption of $\Gamma=1.8$; the median value of the \textit{Swift}/BAT
70-month AGN sample \citep{ric16}.
On the other hand, \cite{mar04} account for variations in AGN SEDs
to obtain the bolometric correction with AGN luminosity.
They assume a varying relation between optical/UR and X-ray luminosity,
which is called a luminosity-dependent bolometric correction.
This gives the larger bolometric correction than the constant one
in higher AGN luminosity end, which would make average $\lbol$ larger,
and $C_{\rm T}$ smaller.

The right panel of Figure~\ref{fig:CFvsLbol_M04} shows the same plot of Figure~\ref{fig:CFvsLbol},
but for the one using the luminosity-dependent bolometric correction of \citep{mar04}.
As expected from the luminosity-dependent bolometric correction, 
the distribution is slightly shifted into the right and bottom direction in the Figure.
Actually, the median values of AGN bolometric luminosity and $\cfdust$
changes from $(\log \lbol, \cfdust)=(44.65, 0.46)$ to 
$(\log L_{\rm bol}^{(\rm AGN; M04)}, \cfdust)=(44.79, 0.39)$.

The figure clearly holds the trend of $\cfgasdust \geq \cfdust$
over the entire AGN luminosity range.
On the other hand, the slight decline of the $\cfdust$ at the lowest AGN bolometric
luminosity bin disappears in Figure~\ref{fig:CFvsLbol_M04}. 
This is mainly because of the small
statistics in the lowest luminosity bin and some sources are shifted into
the higher luminosity bin because of the larger bolometric correction by \cite{mar04}.

%------------------------------------------Figure~CFvsLbol_diffParam-------------------------------------%
\begin{figure*}
\begin{center}
\includegraphics[width=0.33\linewidth]
{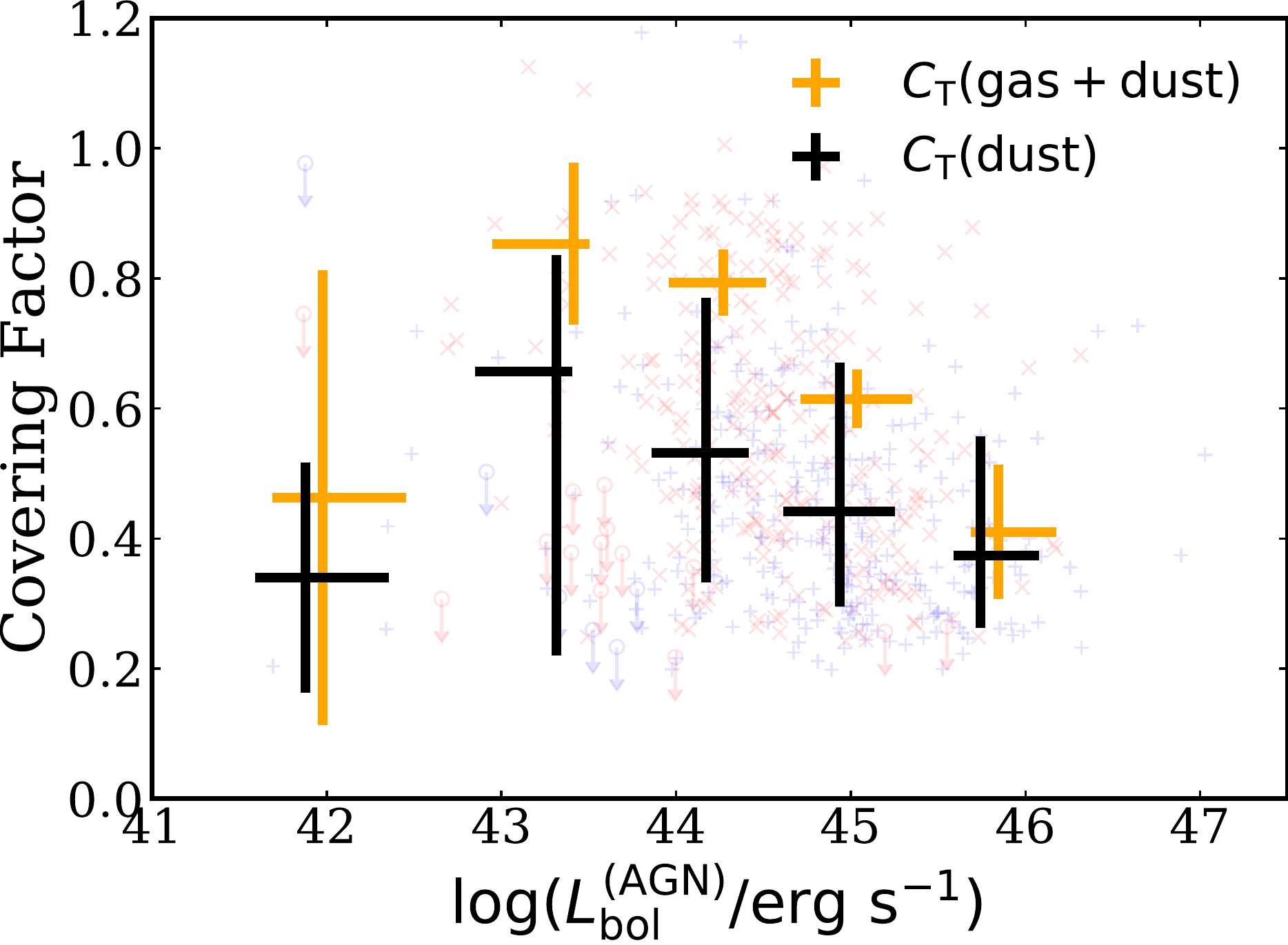}~
\includegraphics[width=0.33\linewidth]
{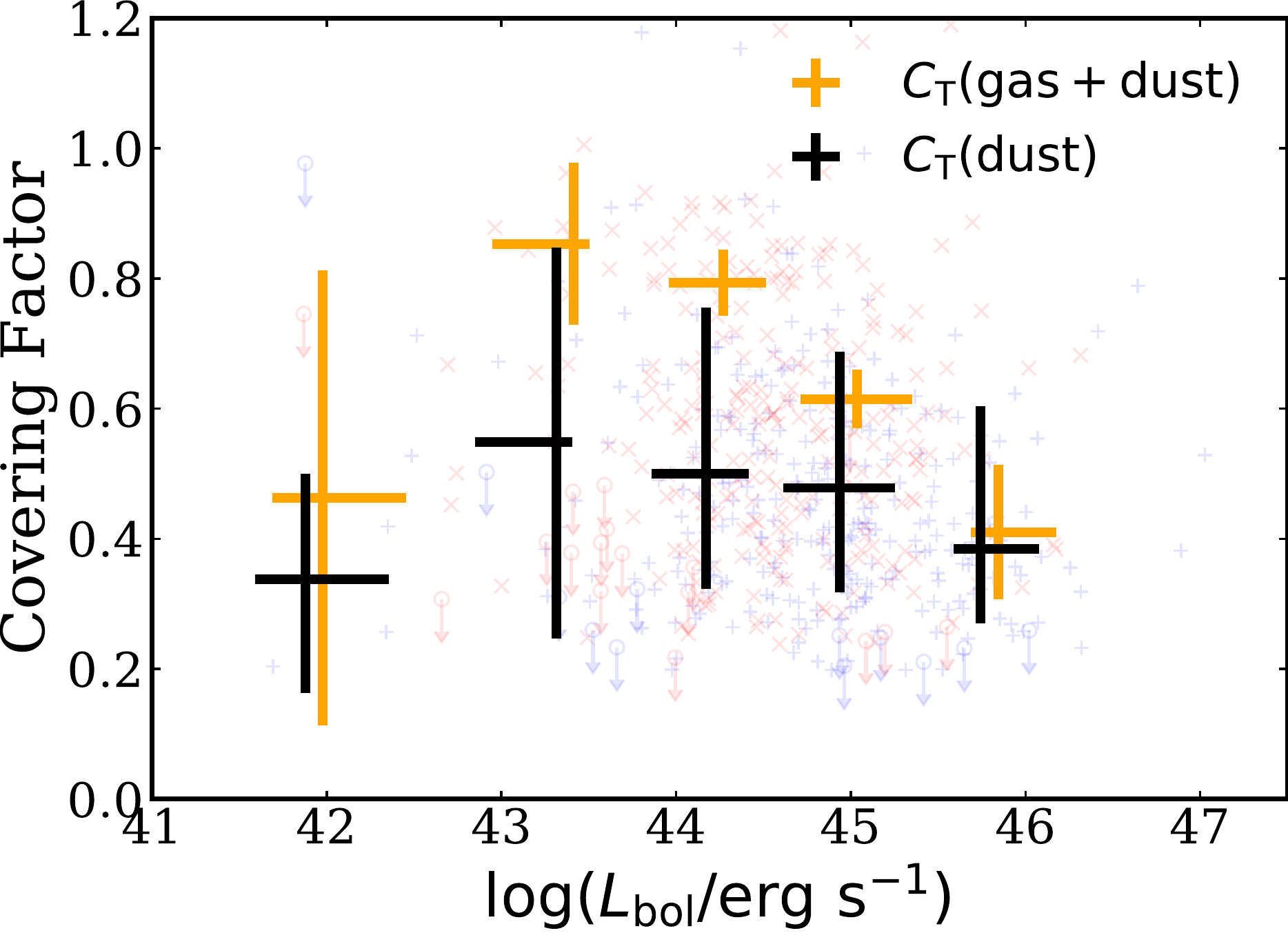}~
\includegraphics[width=0.33\linewidth]
{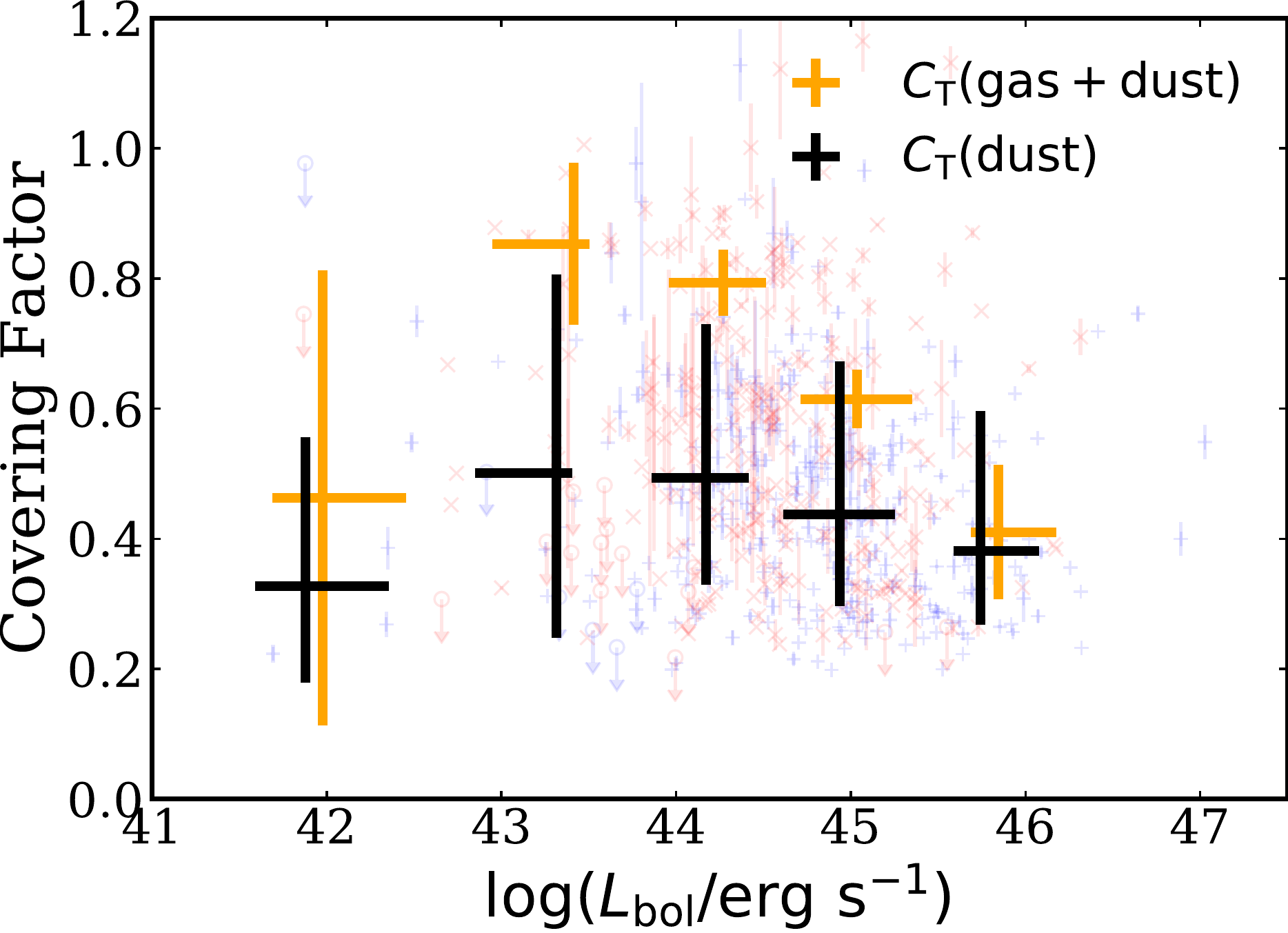}~
\caption{ 
Same as Figure~\ref{fig:CFvsLbol}, but using the different set of free-parameters
for (Left) addition of the dust extinction, and (Middle) fixed the power-law index $\alpha_1 = 1.8$. (Right) same as Figure~\ref{fig:CFvsLbol}, but using the averaged $\cfdust$ of the SB templates as discussed in Appendix~\ref{sec:addSB}.
}\label{fig:CFvsLbol_diffP}
\end{center}
\end{figure*}
%------------------------------------------Figure~CFvsLbol_diffParam-------------------------------------%

\subsection{Dependence of Additional Torus Parameters}\label{sec:addparam}
We here discuss how the dust covering factor changes when we change
the set of the torus parameters. In this study we have only considered
the spectral power-law index ($\alpha_1$) at $\lambda < 19$~$\mu$m
for high luminosity end with $\log L_{14-150} > 44$,
but not considered the dust extinction for obscured AGN, which
could be one of the most significant parameters shaping the torus SEDs.
The left panel of Figure~\ref{fig:CFvsLbol_diffP} shows the $\cfdust$ 
as a function of $\lbol$ after addition of the dust extinction for obscured AGN
using the absorption profile of \cite{dra03} \citep[see also ][]{mul11}.
$\cfdust$ becomes slightly larger, but the overall sense does not change.
The middle panel shows the same plot using a fixed power-law index $\alpha_1=1.8$
for all sources without the dust extinction. The $\cfdust$ shows a relatively flatter
distribution compared to Figure~\ref{fig:CFvsLbol}, but the overall trend
of $\cfdust < \cfgasdust$ still holds.

\subsection{Dependence of Other SB Templates}\label{sec:addSB}
In this study, we used the best SB template based on the
lowest $\chi^2$ value as discussed in Section~\ref{sec:SEDfitting}.
However, the other SB templates sometimes show similar quality
fitting results with small $\Delta \chi^2$ between the best one and the other.
Therefore, we here investigate how the result could be affected by using such
different SB templates. We consider here that the fitting result is indistinguishable
if the $\Delta \chi^2$ between the best fitting SB template and
the other SB ones are smaller than the $\chi^2_{\rm max}$, which
is the maximum allowed $\chi^2$ corresponding the $p$-value$=0.05$ of $\chi^2$ distributions with the degree of freedom for each source.
If each source has indistinguishable SB templates, we then
measure the averaged $\cfdust$ and the standard deviation $\Delta \cfdust$.
The right panel of Figure~\ref{fig:CFvsLbol_diffP} shows the result using the averaged
$\cfdust$ here, and the binned values of $\cfdust$ becomes slightly smaller compared to the original ones.

\section{Full List of SEDs}\label{sec:fullSED}
The full list of SEDs with the fitting results is available from the link here\footnote{\url{http://www.kusastro.kyoto-u.ac.jp/~ichikawa/ms_KI_20180307_fullSED.pdf}}.

%% Tables may also be prepared as separate files. See the accompanying
%% sample file table.tex for an example of an external table file.
%% To include an external file in your main document, use the \input
%% command. Uncomment the line below to include table.tex in this
%% sample file. (Note that you will need to comment out the \documentclass,
%% \begin{document}, and \end{document} commands from table.tex if you want
%% to include it in this document.)


%% The following command ends your manuscript. LaTeX will ignore any text

\listofchanges

\end{document}